\documentclass[aps,11pt,preprintnumbers,nofootinbib,floatfix]{revtex4}

\usepackage{graphicx}
\usepackage{epsfig}
\usepackage{dcolumn}
\usepackage{subfigure}
\usepackage{multirow}

\def\be{\begin{equation}}
\def\ee{\end{equation}}
\def\bea{\begin{eqnarray}}
\def\eea{\end{eqnarray}}

\begin{document}

\title{Flipped $U(1)$ extended Standard Model and Majorana dark matter}


\author{Cao H. Nam}
\email{nam.caohoang@phenikaa-uni.edu.vn}  
\affiliation{Phenikaa Institute for Advanced Study and Faculty of Basic Science, Phenikaa University, To Huu, Yen Nghia, Ha Dong, Hanoi 100000, Vietnam}
\date{\today}

\begin{abstract}%
We propose a general flavor-independent extension of the Standard Model (SM) with the minimal particle content, based on the symmetry $SU(3)_C\times SU(2)_L\times U(1)_{Y'}\times U(1)_X\times Z_2$. In this scenario, the charge operator is identified in terms of the charges of two $U(1)$ gauge symmetries. The light neutrino masses are generated via Type-I seesaw mechanism only with two heavy right-handed neutrinos acquiring their Majorana masses through the $U(1)_{Y'}\times U(1)_X$ symmetry breaking. We study various experimental constraints on the parameters of the model and investigate the phenomenology of the right-handed neutrino dark matter (DM) candidate assigned a $Z_2$-odd parity. We find that the most important constraints are the observed DM relic abundance, the current LHC limits, and the ambiguity of the SM neutral gauge boson mass.
\end{abstract}

\maketitle
\section{Introduction}

Although the Standard Model (SM) is successful in explaining most of the observed elementary particle phenomena, there are at least two evidences which hint new physics beyond the SM, such as nonzero but tiny neutrino masses and the dark matter (DM). A simple extension of the SM is to introduce an exotic $U(1)$ gauge symmetry which corresponds to a new short-range neutral gauge boson. Various models have been proposed from the top-down approach \cite{Langacker1981,Rizzo1989,Nam2019a,Nam2020a,Nam2020b} and the bottom-up one
\cite{Davidson1979,Mo-Marshak1980,XHe1991,Baek2001,Appelquist2003,Tait2004,Khalil2008,VPleitez2009,Khalil2010,Heeck2011,Montero2011,WChao2011,
Latosinski2013,MDas2014,Schmitz2014,Altmannshofer2014,SBaek2015,
MaPollard2015,LeeYun2016,WChao2016,ABiswas2016,
Das2017,DelleRose2017,Biswas2017,Asai2017,Singirala2018,Das-Raut2017,TNomura2018,SLee2018,Escudero2018,Nomura2018,
DelleRose2018,Kamada2018,Arcadi2018,Banerjee2018,Camargo2019,Marzo2019,
XHou2019,CWChiang2019,Nam2020c,Choudhury2020,Dong032020,Loi042020,Borah2020}. In the $U(1)$ extensions of the SM, the right-handed neutrinos and even exotic chiral/Dirac fermions are introduced to achieve the anomaly cancellations, generate the masses for the observed neutrinos via the well-known seesaw mechanism, and provide the DM candidates. The DM phenomenology has been investigated in various $U(1)$ extensions of the SM \cite{Belanger2008,Okada2010,Kanemura2011,ZhengYu2012,Okada2012,Lindner2014,Alves2014,Agrawal2014,Alves2015,
Schmitz2015,Duerr2015,Berlin2015,Okada2016,Kaneta2017,SingiralaPatra2018,Chao2017,Borah2017,
Bandyopadhyay2018,Liu2018,Hutauruk2019,Frank2019,Raut2019,Gu2020}. 

In traditional $U(1)$ extensions of the SM, the exotic $U(1)$ charge is usually not related to the charge operator. However, it is natural to consider that the charge operator would be identified in terms of all charges of the present $U(1)$ gauge symmetries. As a result, an extension of the SM in this way should be based on the following symmetry, $\text{GSM}\equiv SU(3)_C\times SU(2)_L\times U(1)_{Y'}\times U(1)_X$, where the symmetry $U(1)_{Y'}$ is in general different to the weak hypercharge symmetry $U(1)_Y$ of the SM, which is called flipped $U(1)$ extended SM. The charge operator in this scenario is identified as follows
\begin{eqnarray}
Q=T_3+Y'+X\label{QO}
\end{eqnarray}
where $Y'$ and $X$ are the $U(1)_{Y'}$ and $U(1)_X$ charges, respectively. This is the motivation for the present work which we construct a general flavor-independent model in a minimal way which is consistent with the experimental constraints as well as incorporates the neutrino masses and a DM candidate.

This work is organized as follows. In Sec. \ref{model}, we present the model construction. In this section, we introduce the field content and charge assignment, and then we study neutrino and neutral gauge sectors and determine the couplings of the neutral gauge bosons to the fermions. In Sec. \ref{constr}, we analyze the experimental constraints on the mass and gauge coupling of the new gauge boson. In Sec. \ref{DM}, we study the DM phenomenology with the third right-handed neutrino stabilized by the $Z_2$ discrete symmetry and hence playing the role of DM candidate. Finally, we conclude in Sec. \ref{conclu}.

\section{\label{model}Model construction}
We consider a model based on the symmetry $\text{GSM}\times Z_2$ with the field content and charge assignment under this symmetry listed on Table \ref{gauge-charge}. Under the $Z_2$ discrete symmetry, only the third right-handed neutrino is odd, whereas other fields take even parity. This would guarantee the stability of the third right-handed neutrino and hence it can play the role of the DM candidate. 

Before proceeding, we pause here to comment on the $Z_2$ discrete symmetry introduced above. We can consider this $Z_2$ discrete symmetry as a reduced symmetry which arises from the high-scale symmetry breaking of a certain $U(1)$. Under that $U(1)$, the charge assignment for the first two right-handed neutrinos is different to that for the third right-handed neutrino, which satisfies the cancellation of all anomalies. (An example of such a charge assignment has been considered for the case of $U(1)_{B-L}$ \cite{VPleitez2009}.) Because of this charge assignment, the reduced $Z_2$ charge of the first two right-handed neutrinos is generally different to that of the third right-handed neutrino. With a consistent charge assignment, it may thus lead to that the first two right-handed neutrinos and the SM particles are even under the reduced $Z_2$ symmetry whereas the third right-handed neutrino is odd.

\begin{table}[!h]
\begin{center}
\begin{tabular}{|c|c|c|c|c|c|}
  \hline
  $ $ &\ \ $SU(3)_C$\ \ & \ \ $SU(2)_L$\ \ & \ \ $U(1)_{Y'}$\ \ & \ \ $U(1)_X$\ \  \ \ &\ \ $Z_2$\ \ \\
  \hline
  \ \ $Q_a$\ \ & $\textbf{3}$  & $\textbf{2}$ & $\frac{1}{6}+z_2-z_1$ & $z_1-z_2$ & $+$ \\
  $u_{aR}$ & $\textbf{3}$  & $\textbf{1}$ & $\frac{2}{3}-z_1$ & \ \ $z_1$ \ \ & $+$ \\
  $d_{aR}$ & $\textbf{3}$  & $\textbf{1}$ & $2z_2-z_1-\frac{1}{3}$ & $z_1-2z_2$ & $+$ \\
  $L_a$ & $\textbf{1}$  & $\textbf{2}$ & $3(z_1-z_2)-\frac{1}{2}$ & $3(z_2-z_1)$ & $+$ \\
  $\nu_{iR}$ & $\textbf{1}$  & $\textbf{1}$ & $3z_1-4z_2$ & $4z_2-3z_1$ & $+$ \\
  $\nu_{3R}$ & $\textbf{1}$  & $\textbf{1}$ & $3z_1-4z_2$ & $4z_2-3z_1$ & $-$ \\
  $e_{aR}$ & $\textbf{1}$  & $\textbf{1}$ & $3z_1-2z_2-1$ & $2z_2-3z_1$ & $+$ \\
  $H$ & $\textbf{1}$  & $\textbf{2}$ & $\frac{1}{2}-z_2$ & $z_2$ & $+$ \\
  $\Phi$ & $\textbf{1}$  & $\textbf{1}$ & $2(4z_2-3z_1)$ & $2(3z_1-4z_2)$ & $+$ \\
  \hline
\end{tabular}
\caption{The charge assignement for the fermions and scalars under the symmetry $\text{GSM}\times Z_2$ where $z_1$ and $z_2$ are the free parameters with $a=1,2,3$ and $i=1,2$.}\label{gauge-charge}
\end{center}
\end{table}
Up to the gauge fixing and ghost terms, the kinetic terms of the gauge fields, and the terms related to the $SU(3)_C$ sector, the most total Lagrangian is given by
\begin{eqnarray}
\mathcal{L}&\supset&\sum_{f}\bar{f}i\gamma^\mu D_\mu f+\left(D_\mu H\right)^\dagger(D^\mu H)+\left(D_\mu\Phi\right)^\dagger(D^\mu\Phi)-V(H,\Phi)+\mathcal{L}_{\text{Y}},\label{tot-Lag}
\end{eqnarray}
where $f$ refers to the chiral fermions of our model, $D_\mu=\partial_\mu+ig\frac{\sigma_i}{2}W_{i\mu}+ig_1Y'B_\mu+ig_2X X_{\mu}$ with $\{g,g_1,g_2\}$ to be the gauge couplings corresponding to $\{SU(2)_L, U(1)_{Y'},U(1)_X\}$. The scalar potential $V(H,\Phi)$ takes the following form
\begin{eqnarray}
  V(H,\Phi)=\mu^2_1H^\dagger H+\lambda_1(H^\dagger H)^2+\mu^2_2\Phi^\dagger\Phi+\lambda_2(\Phi^\dagger\Phi)^2+\lambda_3(H^\dagger H)(\Phi^\dagger\Phi),
\end{eqnarray}
where in order to guarantee the potential bounded from below, the coupling constants must satisfy the condition as
\begin{eqnarray}
\lambda_1,\lambda_2>0,\ \ \ \ \lambda_3>-2\sqrt{\lambda_1\lambda_2}.
\end{eqnarray}
The Yukawa interactions are given by
\begin{eqnarray}
-\mathcal{L}_{\text{Y}}&=&\sum^3_{a,b=1}\left(h^e_{ab}\bar{L}_aHe_{bR}+h^d_{ab}\bar{Q}_aHd_{bR}+h^u_{ab}\bar{Q}_a\widetilde{H}u_{bR}+h^M_{ab}\bar{\nu}^C_{aR}\nu_{bR}\Phi\right)\nonumber\\
&&+\sum^3_{a=1}\sum^2_{i=1}h^\nu_{ai}\bar{L}_a\widetilde{H}\nu_{iR}+\textrm{H.c},
\end{eqnarray}
with $\widetilde{H}=i\sigma_2H^*$ and without loss of generality the matrix $h^M $ to be diagonal for simplicity. Note that, the term $\bar{L}_a\widetilde{H}\nu_{3R}$ is forbidden by the $Z_2$ charge assignment.

Because the weak hypercharge symmetry $U(1)_Y$ is identified as the remaining symmetry of the spontaneously $U(1)_{Y'}\times U(1)_X$ symmetry breaking, one can find the relation between the gauge coupling $g'$ of the symmetry $U(1)_Y$ in terms of $g_1$ and $g_2$ as
\begin{eqnarray}
g'=\frac{g_1g_2}{\sqrt{g^2_1+g^2_2}}.
\end{eqnarray}

We discuss the assignment of the $U(1)_{Y'}$ and $U(1)_X$ charges from the conditions of the anomaly cancellation and gauge invariance of the Yukawa couplings. The nontrivial anomalies in our model are listed as follows: $[SU(3)_C]^2U(1)_{Y'}$, $[SU(2)_L]^2U(1)_{Y'}$, $[U(1)_{Y'}]^2U(1)_X$, $[U(1)_{Y'}]^3$, $[\text{Gravity}]^2U(1)_{Y'}$, $[SU(3)_C]^2U(1)_X$, $[SU(2)_L]^2U(1)_X$, $[U(1)_X]^2U(1)_{Y'}$, $[U(1)_X]^3$, and $[\text{Gravity}]^2U(1)_X$. We can check that because of the relation $Y=Y'+X$ the anomaly cancellation conditions associated with the last five anomalies are automatically satisfied if the first five anomalies vanish, and vice versa. Therefore, we can write the anomaly cancellation conditions as
\begin{eqnarray}
&&[SU(3)_C]^2U(1)_X: \ \ 2X_Q-X_{u_R}-X_{d_R}=0,\nonumber\\
&&[SU(2)_L]^2U(1)_X: \ \ 3X_Q+X_L=0,\nonumber\\
&&[U(1)_X]^2U(1)_{Y'}:  \ \ 2\left(3X^2_QY'_Q+X^2_LY'_L\right)-3(X^2_{u_R}Y'_{u_R}+X^2_{d_R}Y'_{d_R})-X^2_{e_R}Y'_{e_R}-X^2_{\nu_R}Y'_{\nu_R}=0,\nonumber\\
&&[U(1)_X]^3:  \ \ 2\left(3X^3_Q+X^3_L\right)-3(X^3_{u_R}+X^3_{d_R})-X^3_{e_R}-X^3_{\nu_R}=0,\nonumber\\
&&[\text{Gravity}]^2U(1)_X: \ \ 2\left(3X_Q+X_L\right)-3\left(X_{u_R}+X_{d_R}\right)-X_{e_R}-X_{\nu_R}=0.\label{anoca}
\end{eqnarray}
In addition, the gauge invariance of the Yukawa couplings leads to
\begin{eqnarray}
-X_Q-X_H+X_{u_R}&=&-X_Q+X_H+X_{d_R}=0,\nonumber\\
-X_L-X_H+X_{\nu_R}&=&-X_L+X_H+X_{e_R}=0,\nonumber\\
2X_{\nu_R}+X_\Phi&=&0.\label{Yuconds}
\end{eqnarray}
Note that, the similar conditions for the $U(1)_{Y'}$ charges are automatically satisfied because of the relation $Y=Y'+X$. With ({\ref{Yuconds}), the first and fifth lines in (\ref{anoca}) are automatically satisfied. Whereas, the remaining conditions in (\ref{anoca}) lead to unique solution as
\begin{eqnarray}
X_{\nu_R}=4X_H-3X_{u_R}.
\end{eqnarray}
Then, we can express the $U(1)_{Y'}$ and $U(1)_X$ charges of the fields in terms of two independent parameters which are the $U(1)_X$ charges of the scalar doublet $H$ and right-handed up-type quark, as seen in Table \ref{gauge-charge}.

The scalar fields develop their vacuum expectation values (VEVs) given as, $\langle H\rangle=(\begin{array}{cc} 0 & v/\sqrt{2}\end{array})$ and $\langle\Phi\rangle=v'/\sqrt{2}$. The VEV of $\Phi$ would break the symmetry $U(1)_{Y'}\times U(1)_X$ down to $U(1)_Y$, and the VEV of $H$ would break the electroweak symmetry $SU(2)_L\times U(1)_Y$ down to $U(1)_{\text{em}}$. After the spontaneous symmetry breaking, the mass matrices for the observed light neutrinos $\nu'_L$ and the heavy neutrinos ${\nu'_R}$ are given by
\begin{eqnarray}
m_{\nu'_L}\simeq-U_{_{MNS}}M_DM^{-1}_MM^T_DU^\dagger_{_{MNS}},\ \ \ \ \ \ m_{\nu'_R}\simeq M_M,\label{neutmasses}
\end{eqnarray}
where $\left[M_D\right]_{ai}=\frac{v}{\sqrt{2}}h^\nu_{ai}$ and $\left[M_D\right]_{a3}=0$ (with $a=1,2,3$ and $i=1,2$), $\left[M_M\right]_{ab}=\sqrt{2}v'h^M_{ab}$, and $U_{_{MNS}}$ is the Maki-Nakagawa-Sakata (MNS) matrix determined by the current neutrino oscillation data \cite{Tanabashi2018}. Because of $\text{det}\left(m_{\nu'_L}\right)=0$ the model predicts one massless light neutrino eigenstate, which is still consistent with the neutrino oscillation data \cite{Frampton2002}. The mass eigenstates are related to the flavor states as
\begin{eqnarray}
\left(\begin{array}{c}
\nu'^C_L \\
\nu'_R \\
\end{array}
\right) &\approx&\left(
\begin{array}{cc}
U_{_{MNS}} & -\delta^\dagger \\
\delta & 1\\
\end{array}\right)\left(\begin{array}{c}
\nu^C_L \\
\nu_R \\
\end{array}
\right),
\end{eqnarray}
where $\delta=M^{-1}_MM^T_D$. Because the mixing parameter $\delta$ is extremely small due to the sub-eV neutrino mass scale, we have the approximation, $\nu'_L\approx U_{_{MNS}}\nu_L$ and $\nu'_R\approx\nu_R$.

The first expression of Eq. (\ref{neutmasses}) indicates that the light SM neutrino masses are generated via Type-I seesaw mechanism with two right-handed neutrinos. If the Dirac-Yukawa coupling constants $h^\nu_{ai}$ are not too small, then very large masses of the right-handed neutrinos are necessary to generate the sub-eV mass scale for the light SM neutrinos. For instance, with the Dirac-Yukawa coupling constants $h^\nu_{ai}$ to be of order unity (the order of the Yukawa coupling constant of the top quark), the right-handed neutrino masses are required to be of order $\mathcal{O}(10^{14})$ GeV. The very large mass scale of the right-handed neutrinos is well motivated from the grand unification models. However, no any experimental evidence indicates that the right-handed neutrino masses should be of order the grand unification or around. The low-scale Type-I seesaw scenario with the right-handed neutrino masses to be of order TeV is very attractive due to the potential discovery of new physics at the LHC as well as the signature modifications on the SM phenomenology. In this seesaw scenario, the Dirac-Yukawa coupling constants $h^\nu_{ai}$ have to be very small, which are of order $\mathcal{O}(10^{-6})$.

The mass matrix of the neutral gauge bosons is given in the basis $(W_{3\mu}, B_\mu, X_\mu)$ as
\begin{eqnarray}
M^2_{\text{NGB}}=\left(%
\begin{array}{ccc}
  \frac{g^2v^2}{4} & \star & \star \\
  -\frac{gg_1(1-2z_2)v^2}{4} & \frac{g^2_1}{4}\left[(1-2z_2)^2v^2+16(4z_2-3z_1)^2{v'}^2\right] & \star \\
  -\frac{gg_2z_2v^2}{2} & \frac{g_1g_2}{2}\left[(1-2z_2)z_2v^2-8(4z_2-3z_1)^2v'^2\right] & g^2[z^2_2v^2+4(4z_2-3z_1)^2v'^2] \\
\end{array}%
\right).\nonumber\\
\end{eqnarray} 
The corresponding mass eigenvalues are obtained as, $\text{Diag}(M^2_Z,M^2_{Z'},0)=VM^2_{\text{NGB}}V^T$,  where
\begin{eqnarray}
M^2_{Z}&=&\frac{1}{4}(g^2+g'^2)v^2-\frac{g^2+g'^2}{64(3z_1-4z_2)^2}\left[\frac{2(1+t^2)z_2-1}{1+t^2}\right]^2\frac{v^4}{v'^2}+\mathcal{O}(v^4/v'^4),\nonumber\\
M^2_{Z'}&=&4(g^2_1+g^2_2)(3z_1-4z_2)^2v'^2+\frac{(g^2_1+g^2_2)}{4}\left[\frac{2(1+t^2)z_2-1}{1+t^2}\right]^2v^2+\mathcal{O}(v^2/v'^2),\label{Zpmass}
\end{eqnarray}
with $t\equiv g_2/g_1$ and the diagonalizing matrix $V$ is given by
\begin{eqnarray}
V=\left(%
\begin{array}{ccc}
 c_\beta  & -s_\beta &  0\\
 s_\beta & c_\beta & 0 \\
 0 & 0 & 1 \\ 
\end{array}%
\right)\left(%
\begin{array}{ccc}
 c_W  & 0 & -s_W \\
 0 & 1 & 0 \\
 s_W & 0 & c_W \\ 
\end{array}%
\right)\left(%
\begin{array}{ccc}
 1  & 0 &  0\\
 0 & \frac{g_1}{\sqrt{g^2_1+g^2_2}} & -\frac{g_2}{\sqrt{g^2_1+g^2_2}} \\
 0 & \frac{g_2}{\sqrt{g^2_1+g^2_2}} & \frac{g_1}{\sqrt{g^2_1+g^2_2}} \\ 
\end{array}%
\right),
\end{eqnarray}
with $\theta_W$ to be the Weinberg angle and the mixing angle $\beta$ given by
\begin{eqnarray}
\tan2\beta&=&\frac{t\left[2(1+t^2)z_2-1\right]}{8s_W(3z_1-4z_2)^2(1+t^2)^2}\frac{v^2}{v'^2}+\mathcal{O}(v^4/v'^4),\nonumber\\
&=&2s_W\frac{\left[2(1+t^2)z_2-1\right]}{t}\frac{M^2_{Z}}{M^2_{Z'}}+\mathcal{O}(M^4_{Z}/M^4_{Z'}).\label{ZZpmixing}
\end{eqnarray}
The physical states are related to $(W_{3\mu}, B_\mu, X_\mu)$ as, $(Z_{\mu},Z'_{\mu},A_\mu)^T=V(W_{3\mu}, B_\mu, X_\mu)^T$.

The couplings of the neutral gauge bosons to the SM chiral fermions and the heavy neutrinos $\nu'_{aR} $ are given as
\begin{eqnarray}
\mathcal{L}_{NC}&=&-\sum_{f_i}\left(C^Z_{f_i}\bar{f}_i\gamma^\mu f_iZ_{\mu}+C^{Z'}_{f_i}\bar{f}_i\gamma^\mu f_iZ'_{\mu}\right),
\end{eqnarray}
where $i=L,R$ and the coupling factors read
\begin{eqnarray}
C^{Z}_{f_i}&=&\frac{gc_\beta}{c_W}\left[T_3(f_i)-Q_{f}s^2_W\right]-\frac{gt_Ws_\beta}{t}\left(Y'_{f_i}-t^2X_{f_i}\right),\nonumber\\
C^{Z'}_{f_i}&=&\frac{gs_\beta}{c_W}\left[T_3(f_i)-Q_{f}s^2_W\right]+\frac{gt_Wc_\beta}{t}\left(Y'_{f_i}-t^2X_{f_i}\right).
\end{eqnarray}
More explicitly, we show the $Z$ and $Z'$ couplings to the fermions in Tables \ref{Z} and \ref{Zp}.
\begin{table}[!htp]
\centering
\begin{tabular}{ccc}
\hline\hline
$f$ & $C^{Z}_{V,f}=(C^{Z}_{f_L}+C^{Z}_{f_R})/2$ & $C^{Z}_{A,f}=(C^{Z}_{f_R}-C^{Z}_{f_L})/2$\\ 
\hline
$\ \ \nu'_{1L},\nu'_{2L},\nu'_{3L} \ \ $ & $\ \ \ \ \frac{gc_\beta}{4c_W}-gt_Ws_\beta\frac{6(z_1-z_2)(1+t^2)-1}{4t} \ \ \ \ $ & $\ \ \ \ -\frac{gc_\beta}{4c_W}+gt_Ws_\beta\frac{6(z_1-z_2)(1+t^2)-1}{4t}$ \\
$\ \ \nu'_{1R},\nu'_{2R},\nu'_{3R} \ \ $ & $\ \ \ \ -gt_Ws_\beta\frac{(3z_1-4z_2)(1+t^2)}{2t} \ \ \ \ $ & $\ \ \ \ -gt_Ws_\beta\frac{(3z_1-4z_2)(1+t^2)}{2t}$ \\
$\ \ e,\mu,\tau \ \ $ & $\ \ \ \ \frac{gc_\beta}{4c_W}(4s^2_W-1)-gt_Ws_\beta\frac{2(6z_1-5z_2)(1+t^2)-3}{4t} \ \ \ \ $ & $\ \ \ \ \frac{gc_\beta}{4c_W}-gt_Ws_\beta\frac{2z_2(1+t^2)-1}{4t}$ \\
$\ \ u,c,t \ \ $ & $\ \ \ \ \frac{gc_\beta}{12c_W}(3-8s^2_W)-gt_Ws_\beta\frac{5-6(2z_1-z_2)(1+t^2)}{12t} \ \ \ \ $ & $\ \ \ \ -\frac{gc_\beta}{4c_W}-gt_Ws_\beta\frac{1-2z_2(1+t^2)}{4t}$ \\
$\ \ d,s,b \ \ $ & $\ \ \ \ \frac{gc_\beta}{12c_W}(4s^2_W-3)-gt_Ws_\beta\frac{6(3z_2-2z_1)(1+t^2)-1}{12t} \ \ \ \ $ & $\ \ \ \ \frac{gc_\beta}{4c_W}-gt_Ws_\beta\frac{2z_2(1+t^2)-1}{4t}$ \\
\hline\hline
\end{tabular}
\caption{\label{Z}Vector and axial-vector couplings of $Z$ to the fermions.}  
\end{table}
\begin{table}[!htp]
\centering
\begin{tabular}{ccc}
\hline\hline
$f$ & $C^{Z'}_{V,f}=(C^{Z'}_{f_L}+C^{Z'}_{f_R})/2$ & $C^{Z'}_{A,f}=(C^{Z'}_{f_R}-C^{Z'}_{f_L})/2$\\ 
\hline
$\ \ \nu'_{1L},\nu'_{2L},\nu'_{3L} \ \ $ & $\ \ \ \ \frac{gs_\beta}{4c_W}+gt_Wc_\beta\frac{6(z_1-z_2)(1+t^2)-1}{4t} \ \ \ \ $ & $\ \ \ \ -\frac{gs_\beta}{4c_W}-gt_Wc_\beta\frac{6(z_1-z_2)(1+t^2)-1}{4t}$ \\
$\ \ \nu'_{1R},\nu'_{2R},\nu'_{3R} \ \ $ & $\ \ \ \ gt_Wc_\beta\frac{(3z_1-4z_2)(1+t^2)}{2t} \ \ \ \ $ & $\ \ \ \ gt_Wc_\beta\frac{(3z_1-4z_2)(1+t^2)}{2t}$ \\
$\ \ e,\mu,\tau \ \ $ & $\ \ \ \ \frac{gs_\beta}{4c_W}(4s^2_W-1)+gt_Wc_\beta\frac{2(6z_1-5z_2)(1+t^2)-3}{4t} \ \ \ \ $ & $\ \ \ \ \frac{gs_\beta}{4c_W}+gt_Wc_\beta\frac{2z_2(1+t^2)-1}{4t}$ \\
$\ \ u,c,t \ \ $ & $\ \ \ \ \frac{gs_\beta}{12c_W}(3-8s^2_W)+gt_Wc_\beta\frac{5-6(2z_1-z_2)(1+t^2)}{12t} \ \ \ \ $ & $\ \ \ \ -\frac{gs_\beta}{4c_W}+gt_Wc_\beta\frac{1-2z_2(1+t^2)}{4t}$ \\
$\ \ d,s,b \ \ $ & $\ \ \ \ \frac{gs_\beta}{12c_W}(4s^2_W-3)+gt_Wc_\beta\frac{6(3z_2-2z_1)(1+t^2)-1}{12t} \ \ \ \ $ & $\ \ \ \ \frac{gs_\beta}{4c_W}+gt_Wc_\beta\frac{2z_2(1+t^2)-1}{4t}$\\
\hline\hline
\end{tabular}
\caption{\label{Zp}Vector and axial-vector couplings of $Z'$ to the fermions.}  
\end{table}
In Fig. \ref{couplings}, we show the behavior of the $Z'$ coupling strength $\sqrt{(C^{Z'}_{V,f})^2+(C^{Z'}_{A,f})^2}$ to the fermion $f$ as a function of the ratio $g_2/g_1$ for various values of $z_{1,2}$.
\begin{figure}[t]
 \centering
\begin{tabular}{cc}
\includegraphics[width=0.47 \textwidth]{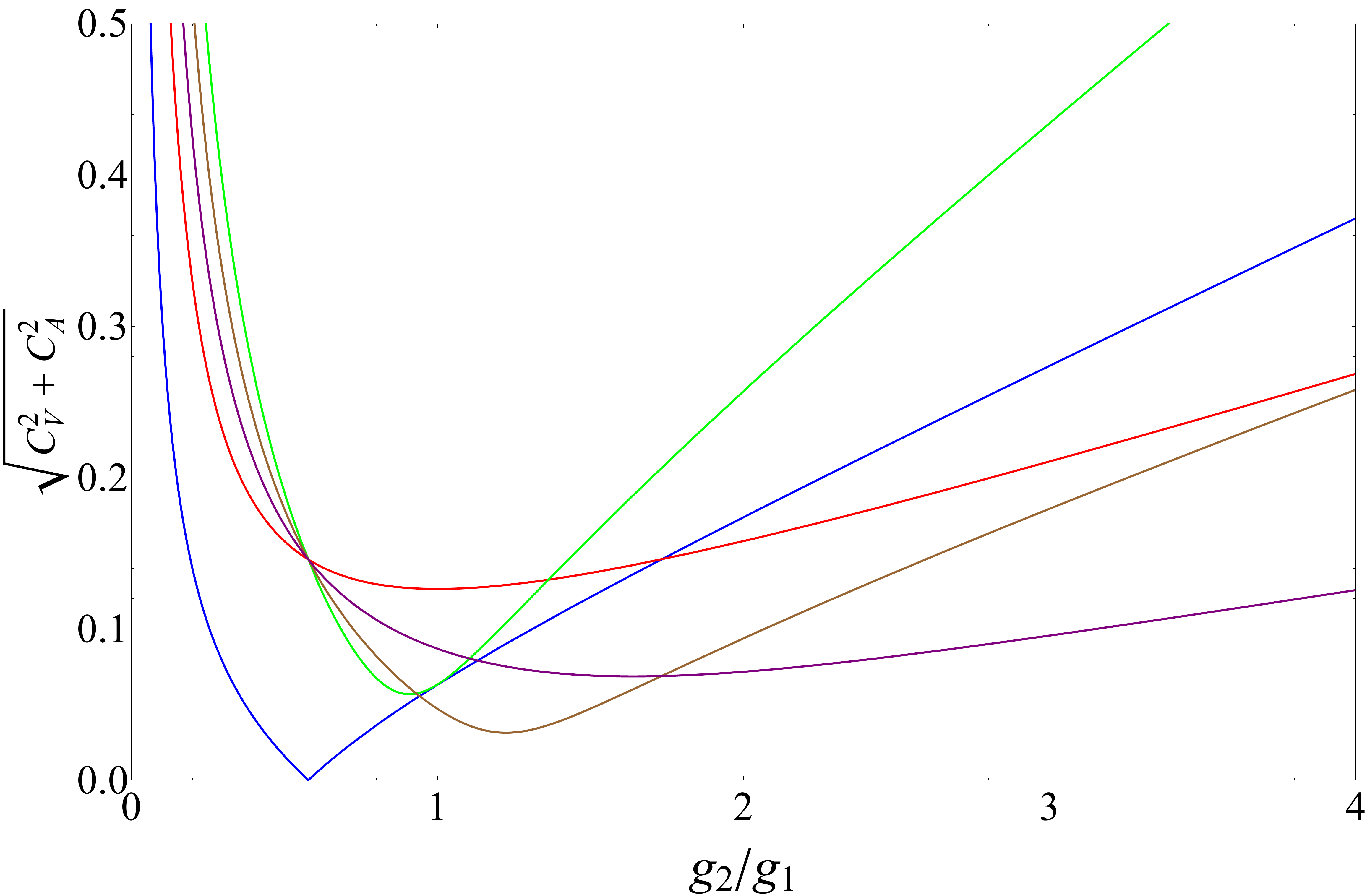}
\hspace*{0.05\textwidth}
\includegraphics[width=0.47 \textwidth]{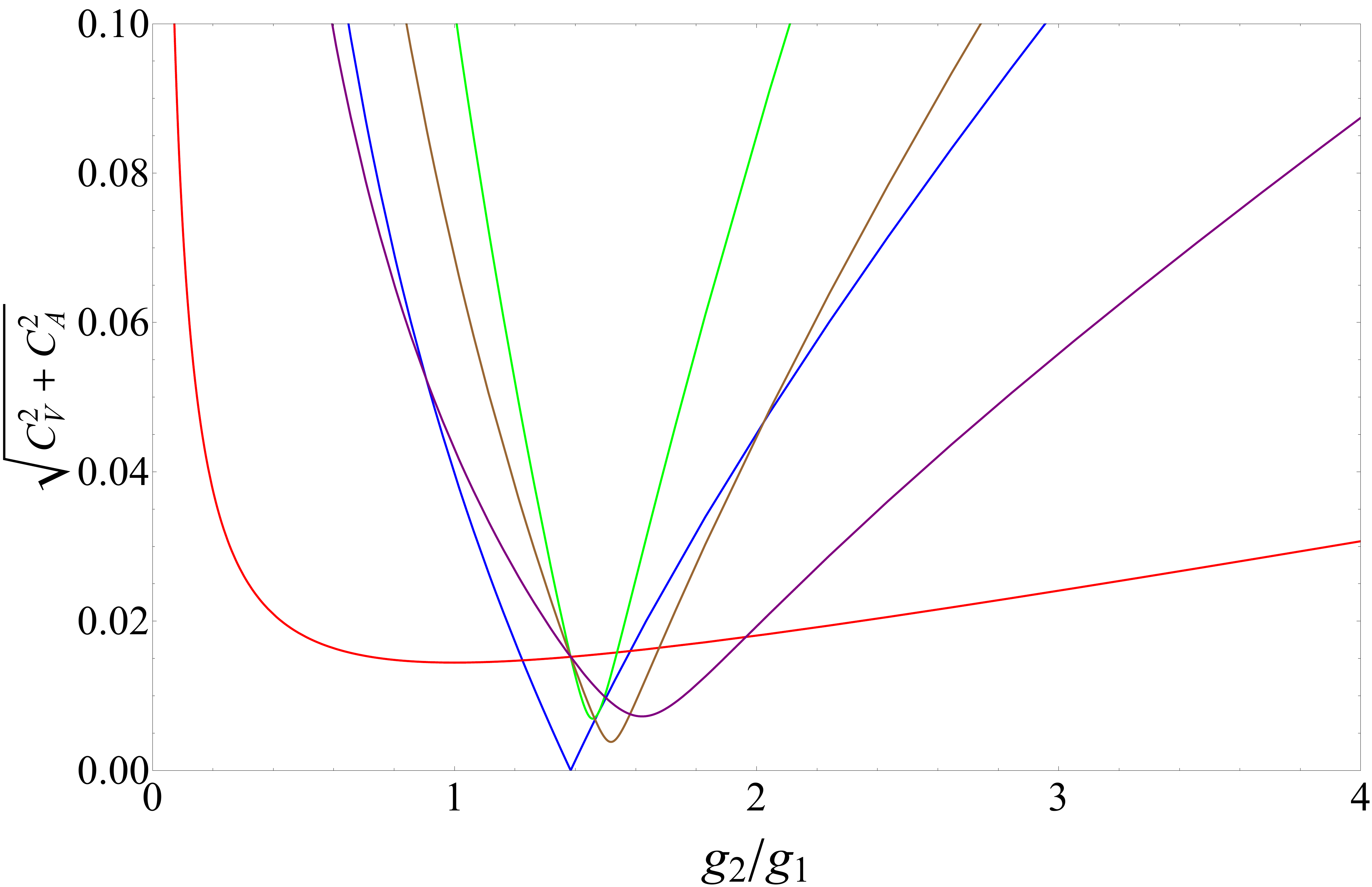}\\
\includegraphics[width=0.47 \textwidth]{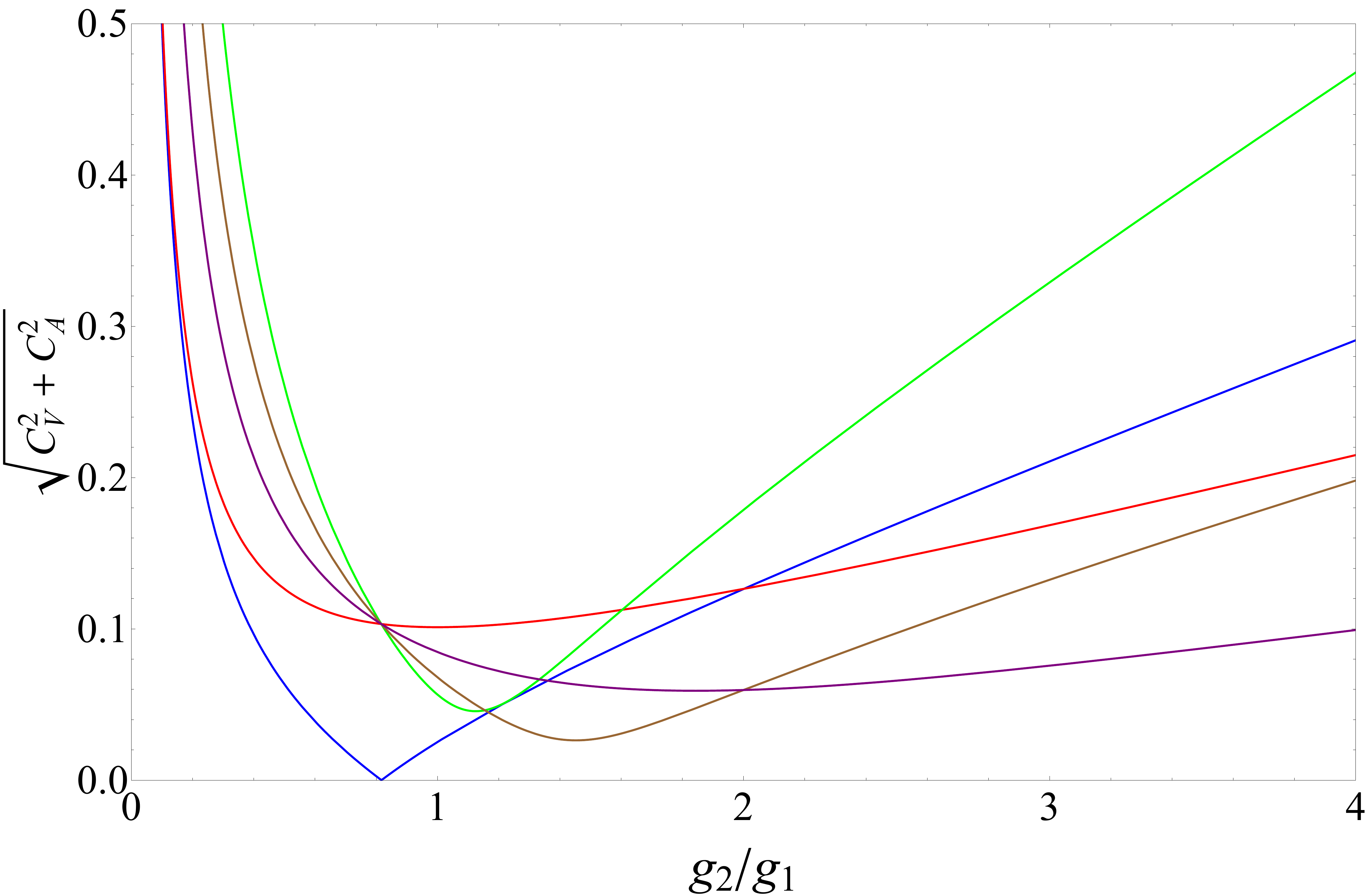}
\end{tabular}
  \caption{The $Z'$ coupling strength $\sqrt{(C^{Z'}_{V,f})^2+(C^{Z'}_{A,f})^2}$ as a function of the ratio $g_2/g_1$ for several values of $z_1$ and $z_2$. The blue, red, green, brown, and purple curves corresponding to $\nu'_L$, $\nu'_R$, $e$, $u$, and $d$, respectively. Top-left panel: $z_1=1/4$ and $z_2=1/8$. Top-right panel: $z_1=1/5$ and $z_2=1/7$. Bottom panel: $z_1=1/5$ and $z_2=1/10$.}\label{couplings}
\end{figure}
From this figure, we see that the $Z'$ coupling strength $\sqrt{(C^{Z'}_{V,f})^2+(C^{Z'}_{A,f})^2}$ should first decrease until a minimum and then increase as increasing the parameter $t$. On the other hand, there exists a certain value of $t$ with given $z_{1,2}$ near which the $Z'$ coupling strength is relatively small but becomes larger as $t$ going far from this value of $t$.

In summary, from Eqs. (\ref{Zpmass}) $\&$ (\ref{ZZpmixing}) and Table \ref{Zp}, the properties of the new gauge boson $Z'$ in our model are primarily described by four independent parameters which are $t$, $z_{1,2}$, and VEV $v'$ of the scalar field $\Phi$. One can not use the redefinition of the couplings $g_{1,2}$ or the charges to define one of the parameters $t$ and $z_{1,2}$ to be $1$.

Note that, the electromagnetic couplings and the charged currents of the quarks do not get modified in our model, whereas the charged currents of the leptons is given by $
\mathcal{L}_{CC}=-\frac{g}{\sqrt{2}}\bar{l}_L\gamma^\mu U^\dagger_{MNS}\nu'_LW^-_\mu+\textrm{H.c.}$.

\section{\label{constr}Constraints}
In this section, we use various current experimental constraints to impose the bounds on the mass of the new gauge boson $Z'$ which are in general the functions of $t$ and $z_{1,2}$.

First, we impose the perturbativity conditions on the gauge couplings as
\begin{eqnarray}
g\leq\sqrt{4\pi},
\end{eqnarray}
for the $SU(2)_L$ and
\begin{eqnarray}
|Y'_{Q,u_R,d_R,L,e_R,\nu_R,H,\Phi}|g_1\leq\sqrt{4\pi},\nonumber\\
|X_{Q,u_R,d_R,L,e_R,\nu_R,H,\Phi}|g_2\leq\sqrt{4\pi},
\end{eqnarray}
for the $U(1)_{Y'}$ and $U(1)_X$, respectively. In addition, the perturbative condition on the Higgs couplings is given as \cite{Huffel1981,Duerr2016}
\begin{eqnarray}
|\lambda_3|\leq8\pi,\ \ \ \ 3(\lambda_1+\lambda_2)\pm\sqrt{\lambda^2_3+9(\lambda_1+\lambda_2)^2}\leq8\pi.
\end{eqnarray}

The precise measurement of the mass of the SM gauge boson $Z$ is given by $M_Z=91.1876\pm0.0021$ GeV \cite{Tanabashi2018}, hence we require the following constraint
\begin{eqnarray}
|\Delta M_Z|\simeq\frac{s^2_W}{2}\left[\frac{2(1+t^2)z_2-1}{t}\right]^2\frac{M^3_{Z}}{M^2_{Z'}}\lesssim0.0021\ \ \text{GeV}.\label{delMZ}
\end{eqnarray}
This leads to a lower bound as
\begin{eqnarray}
M_{Z'}\gtrsim6.46\times\frac{|2(1+t^2)z_2-1|}{t}\ \ \text{TeV}.\label{Zmasscorr}
\end{eqnarray}

The modification of the mass of the SM gauge boson $Z$ also leads to the correction to $\rho$ parameter and thus would constraint the free parameters of the model. The corresponding correction is given by
\begin{eqnarray}
\Delta\rho&=&\frac{M^2_W}{c^2_WM^2_Z}-\rho_{\text{SM}}\nonumber\\
&\simeq&s^2_W\left[\frac{2(1+t^2)z_2-1}{t}\right]^2\frac{M^2_Z}{M^2_{Z'}},
\end{eqnarray}
where $\rho_{\text{SM}}=1$ is the SM prediction. Using the experimental value $\rho=1.00039\pm0.00019$ \cite{Tanabashi2018}, we require new physics satisfying $\Delta\rho<0.00058$ from which we obtain
\begin{eqnarray}
M_{Z'}\gtrsim1.82\times\frac{|2(1+t^2)z_2-1|}{t} \ \ \text{TeV},
\end{eqnarray}
which is clearly weaker than the bound (\ref{Zmasscorr}).

The mixing between the SM gauge boson $Z$ and new one $Z'$ leads to the corrections in the $Z$ couplings to the SM fermions. Thus, we use the precision measurement of the $Z$ decay width to constrain the relevant corrections. First, we rewrite the $Z$ couplings to the SM chiral fermions $f_i$ ($i=L,R$) as
\begin{eqnarray}
\mathcal{L}_{Z}&=&-\bar{f}_i\gamma^\mu C^{\text{SM}}_{f_i}\left(1+\delta_{f_i}\right)f_iZ_{\mu}
\end{eqnarray}
where $C^{\text{SM}}_{f_L}=g\left[T_3(f_L)-Q_fs^2_W\right]/c_W$ and $C^{\text{SM}}_{f_R}=-\frac{g}{c_W}Q_fs^2_W$ are the SM values for the $Z$ couplings to the left- and right-handed fermions, respectively, and the corrections to the SM predictions of the $Z$ couplings are given by
\begin{eqnarray}
\delta_{f_i}&=&-s^2_W\frac{2(1+t^2)z_2-1}{t^2}\frac{Y'_{f_i}-t^2X_{f_i}}{T_3(f_i)-Q_fs^2_W}\frac{M^2_Z}{M^2_{Z'}}+\mathcal{O}(M^4_{Z}/M^4_{Z'}).
\end{eqnarray}
Then, the $Z$ decay width in our model is given as, $\Gamma_{Z}=\Gamma^{\text{SM}}_{Z}+\Delta\Gamma_{{Z}}$, where $\Gamma^{\text{SM}}_{Z}$ is the SM value and the correction $\Delta\Gamma_{{Z}}$ for the SM prediction reads
\begin{eqnarray}
\Delta\Gamma_{{Z}}&\simeq&\frac{M_Z}{12\pi}\sum_{f_i}N_C(f_i)\left(C^{\text{SM}}_{f_i}\right)^2\delta_{f_i}+\frac{\Delta M_Z}{24\pi}\sum_{f_i}N_C(f_i)\left(C^{\text{SM}}_{f_i}\right)^2,
\end{eqnarray}
where $f_i$ refers to all chiral fermions in the SM and $\Delta M_Z=-|\Delta M_Z|$ with $|\Delta M_Z|$ to be given in (\ref{delMZ}). From the experimental value and the SM prediction for the total $Z$ width as, $\Gamma^{\text{exp}}_Z=2.4952\pm0.0023$ GeV and $\Gamma^{\text{SM}}_Z=2.4942\pm0.0008$ GeV \cite{Tanabashi2018}, we require $|\Delta\Gamma_Z|<0.0041$ GeV which leads to
\begin{eqnarray}
M_{Z'}\gtrsim1.96\times\left|\frac{2(1+t^2)z_2-1}{t^2}\left[(0.26+t^2)z_1+(0.41-0.32t^2)z_2-0.5\right]\right|^{1/2}\ \ \text{TeV}.
\end{eqnarray}

The weak nuclear charge of Cesium has been measured to a precision given by $Q^{\text{exp}}_W(^{133}_{\phantom{k} 55}\text{Cs})=-73.16(29)_{\text{exp}}(20)_{\text{th}}$ \cite{Porsev09-10}
which is in agreement with the SM prediction (including electroweak radiative corrections) as $Q^{\text{th}}_W(^{133}_{\phantom{k} 55}\text{Cs})=-73.16(3)$ \cite{Marciano90-92}. This suggests that the contribution of new physics for the nuclear weak charge of Cesium must satisfy $\left|\Delta Q_W(^{133}_{\phantom{k} 55}\text{Cs})\right|\lesssim0.52$. In our model, the exchange of the new gauge boson $Z'$ leads to an additional contribution for the nuclear weak charge of Cesium, which is given as follows \cite{Diener2012}
\begin{eqnarray}
\Delta Q_W(^{133}_{\phantom{k} 55}\text{Cs})=-16\left(\frac{M_{Z}}{M_{Z'}}\right)^2\left(\frac{c_W}{g}\right)^2C^{Z'}_{A,e}\left[(2Z+N)C^{Z'}_{V,u}+(Z+2N)C^{Z'}_{V,d}\right],
\end{eqnarray}
where $C^{Z'}_{V,f}=(C^{Z'}_{f_L}+C^{Z'}_{f_R})/2$, $C^{Z'}_{A,f}=(C^{Z'}_{f_R}-C^{Z'}_{f_L})/2$, $Z=55$, and $N=78$. Then, one can find a lower bound on the mass of the new gauge boson $Z'$ as
\begin{eqnarray}
M_{Z'}\gtrsim0.95\times\left|\frac{2(1+t^2)z_2-1}{t^2}\left[6.76(1+t^2)z_2-6.57(1+t^2)z_1+1\right]\right|^{1/2}\ \ \text{TeV}.
\end{eqnarray}

At the $e^+e^-$ colliders such as the LEP, the new gauge boson $Z'$ would not be directly produced if it is massive enough. However, the presence of $Z'$ would lead to the deviation from the SM prediction in the scattering processes $e^+e^-\rightarrow\bar{f}f$. Hence, we study the constraint on the contact interactions which are induced by the $Z'$ exchange and described by the following effective Lagrangian
\begin{eqnarray}
\mathcal{L}_{\text{eff}}&=&\frac{1}{1+\delta_{ef}}\frac{1}{M^2_{Z'}}\sum_{i,j=L,R}\eta_{ij}\bar{e}_i\gamma_\mu e_i\bar{f}_j\gamma^\mu f_j,
\end{eqnarray}
where $\delta_{ef}=1(0)$ for $f=e$ ($f\neq e$) and $\eta_{ij}=C^{Z'}_{e_i}C^{Z'}_{f_j}$. From the LEP data \cite{Schael2013}, we impose the constraint on this contact interaction as
\begin{eqnarray}
\frac{2\sqrt{2\pi}M_{Z'}}{\sqrt{\left(C^{Z'}_{e_L}\right)^2+\left(C^{Z'}_{e_R}\right)^2}}\gtrsim24.6(17.8)\ \ \text{TeV},
\end{eqnarray}
for $\eta_{LR},\eta_{RL}>0(<0)$ corresponding to $\left[1+6(z_2-z_1)(1+t^2)\right]\left[1+3(z_2-z_1)+t^2(2z_2-3z_1)\right]>0(<0)$. Then, we obtain a lower bound on the mass of the new gauge boson $Z'$ as
\begin{eqnarray}
M_{Z'}\gtrsim1.75(1.27)\times\frac{I(t,z_1,z_2)^{1/2}}{t} \ \ \text{TeV},
\end{eqnarray}
for $\eta_{LR},\eta_{RL}>0(<0)$ where
\begin{eqnarray}
I(t,z_1,z_2)=\left[\frac{1}{2}+3(z_2-z_1)(1+t^2)\right]^2+\left[1+3(z_2-z_1)+t^2(2z_2-3z_1)\right]^2.
\end{eqnarray}

At the LHC, the new gauge boson $Z'$ can be produced through the Drell-Yan process at which the most significant decay channel is $Z'\rightarrow l^+l^-$ ($l=e,\mu$). The cross section of the process $pp\rightarrow Z'\rightarrow l^+l^-$ at a fixed collider center-of-mass energy $\sqrt{s}$ can be written as
\begin{eqnarray}
\sigma(pp\rightarrow Z'\rightarrow l^+l^-)&=&\sum_{q}\int^s_{4m^2_l}d\hat{s}L_{q\bar{q}}(\hat{s})\frac{\hat{s}}{144\pi}\frac{C_S}{\left(\hat{s}^2-M^2_{Z'}\right)^2+M^2_{Z'}\Gamma^2_{Z'}},\label{totcrrs}
\end{eqnarray}
where $\sqrt{\hat{s}}$ is the invariant mass of the dilepton system, the parton luminosities $L_{q\bar{q}}$ are defined by
\begin{eqnarray}
L_{q\bar{q}}(\hat{s})=\int^1_{\frac{\hat{s}}{s}}\frac{dx}{xs}\left[f_q(x,\hat{s})f_{\bar{q}}\left(\frac{\hat{s}}{xs},\hat{s}\right)+f_q\left(\frac{\hat{s}}{xs},\hat{s}\right)f_{\bar{q}}(x,\hat{s})\right],
\end{eqnarray}
with $f_{q(\bar{q})}(x,\hat{s})$ to be the parton distribution function for the quark $q$ (antiquark $\bar{q}$) evaluated at the scale $\hat{s}$ \cite{Stirling2009}, and $C_S=[(C^{Z'}_{q_L})^2+(C^{Z'}_{q_R})^2][(C^{Z'}_{l_L})^2+(C^{Z'}_{l_R})^2]$. In Eq. (\ref{totcrrs}), the decay width of the new gauge boson $Z'$ into the two-body states is given under the assumption (\ref{asumpt}) as
\begin{eqnarray}
\Gamma_{Z'}=\sum_{f\in\text{SM}}\Gamma(Z'\rightarrow\bar{f}f)+\sum^3_{a=1}\Gamma(Z'\rightarrow\bar{\nu}'_{aR}\nu'_{aR})+\Gamma(Z'\rightarrow W^+W^-)+\Gamma(Z'\rightarrow Zh_1),
\end{eqnarray}
where the partial decay widths are given by
\begin{eqnarray}
\Gamma(Z'\rightarrow\bar{f}f)&=&\frac{N_C(f)M_{Z'}}{24\pi}\sqrt{1-\frac{4m^2_f}{M^2_{Z'}}}\left
\{\left[\left(C^{Z'}_{f_L}\right)^2+\left(C^{Z'}_{f_R}\right)^2\right]\left(1-\frac{m^2_f}{M^2_{Z'}}\right)+6C^{Z'}_{f_L}C^{Z'}_{f_R}\frac{m^2_f}{M^2_{Z'}}\right\},\nonumber\\
\Gamma(Z'\rightarrow\bar{\nu}'_{aR}\nu'_{aR})&=&\frac{M_{Z'}}{24\pi}\left(C^{Z'}_{\nu'_R}\right)^2\left(1-\frac{4m^2_{\nu'_{aR}}}{M^2_{Z'}}\right)^{3/2},\nonumber\\
\Gamma(Z'\rightarrow W^+W^-)&=&(gc_Ws_\beta)^2\frac{M^5_{Z'}}{192\pi M^4_W}\left(1-\frac{4M^2_W}{M^2_{Z'}}\right)^{3/2}\left(1+\frac{20M^2_W}{M^2_{Z'}}+\frac{12M^4_W}{M^4_{Z'}}\right),\nonumber\\
\Gamma(Z'\rightarrow Zh_1)&=&\frac{\left(g_{_{Z'Zh_1}}\right)^2M_{Z'}}{192\pi M^2_{Z}}\left[1-\frac{2m^2_{h_1}-10M^2_{Z}}{M^2_{Z'}}+\frac{(m^2_{h_1}-M^2_{Z})^2}{M^4_{Z'}}\right]\nonumber\\
&&\times\left[1-\frac{2(m^2_{h_1}+M^2_{Z})}{M^2_{Z'}}+\frac{(m^2_{h_1}-M^2_{Z})^2}{M^4_{Z'}}\right]^{1/2},
\end{eqnarray}
where $N_C(f)$ is the color number of the fermion $f$, $h_1$ refers to the SM Higgs, and the mass dimension coupling $g_{_{Z'Zh_1}}$ is given by
\begin{eqnarray}
g_{_{Z'Zh_1}}&\simeq&gM_W\left[\frac{c^2_W}{2}s_{2\beta}-t^2_W(1-2z_2)^2\left(\frac{s_W}{t}c_{2\beta}+\frac{1-t^2s^2_W}{2t^2}s_{2\beta}\right)+4z^2_2t^2_W\left(ts_Wc_{2\beta}-\frac{t^2-s^2_W}{2}s_{2\beta}\right)\right.\nonumber\\ &&\left.+2z_2t_W\left(tc_Wc_{2\beta}+\frac{s_{2\theta_W}}{2}s_{2\beta}\right)+2z_2t^2_W(1-2z_2)\left(\frac{(t^2-1)s_W}{t}c_{2\beta}+(1+s^2_W)s_{2\beta}\right)\right.\nonumber\\
&&\left.-(1-2z_2)t_W\left(\frac{c_W}{t}c_{2\beta}-\frac{s_{2\theta_W}}{2}s_{2\beta}\right)\right],
\end{eqnarray}
with $s_{2\beta}\equiv\sin2\beta$, $c_{2\beta}\equiv\cos2\beta$, and $s_{2\theta_W}\equiv\sin2\theta_W$. Note that, the partial decay widths of $Z'\rightarrow W^+W^-$ and $Z'\rightarrow Zh_1$ are approximately equal together as a result of the Goldstone boson equivalence in the high energy limit. If the decay width of $Z'$ is very narrow, Eq. (\ref{totcrrs}) can be approximated as
\begin{eqnarray}
\sigma(pp\rightarrow Z'\rightarrow l^+l^-)\simeq\frac{\pi}{6}\sum_{q}L_{q\bar{q}}(M^2_{Z'})\left[\left(C^{Z'}_{q_L}\right)^2+\left(C^{Z'}_{q_R}\right)^2\right]{\rm Br}(Z'\rightarrow l^+l^-),
\end{eqnarray}
where ${\rm Br}(Z'\rightarrow l^+l^-)$ is the branching ratio of $Z'$ decaying into the given pair $l^+l^-$.  No evidence for the dilepton resonances has been found in the current LHC data and the $95\%$ confidence level (CL) upper bounds on $\sigma\times BR$ of new neutral gauge boson have been produced using $36.1$ fb$^{-1}$ of proton-proton collision data at $\sqrt{s}=13$ TeV \cite{ATLAS2017}. From these upper bounds,
we can impose the constraint on the $Z'$ gauge boson mass and the gauge coupling ratio $t$ for $z_{1,2}$ kept fixed.

In Fig. \ref{APS}, we show the allowed parameter region in the $t-M_{Z'}$ plane, for several values of $z_1$ and $z_2$, by combining the lower bounds corresponding to the current LHC limits, precision measurement of the $Z$ decay width, the ambiguity of the $Z$ boson mass, the LEP data, and weak nuclear charge of Cesium.
\begin{figure}[t]
 \centering
\begin{tabular}{cc}
\includegraphics[width=0.47 \textwidth]{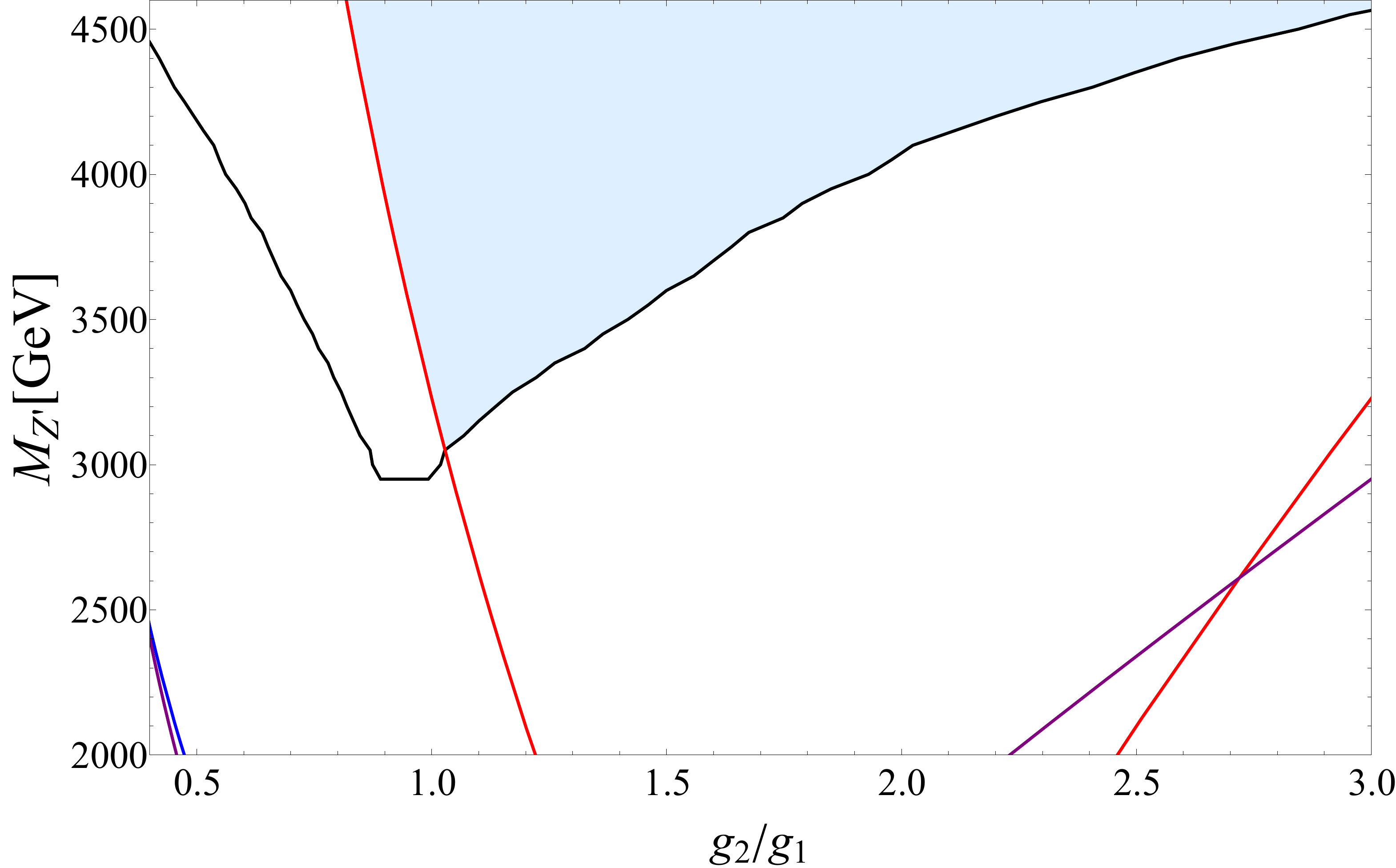}
\hspace*{0.05\textwidth}
\includegraphics[width=0.47 \textwidth]{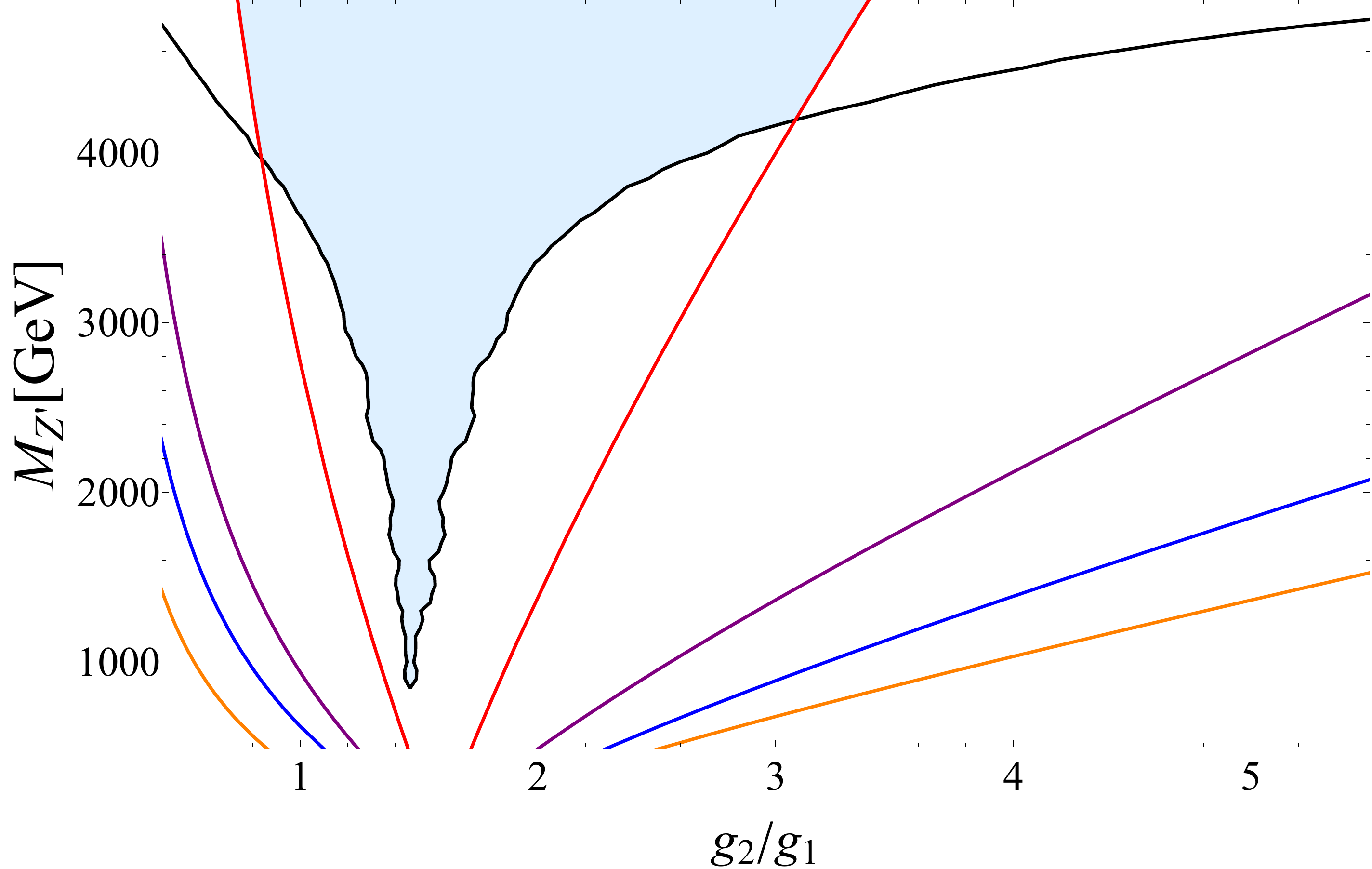}\\
\includegraphics[width=0.47 \textwidth]{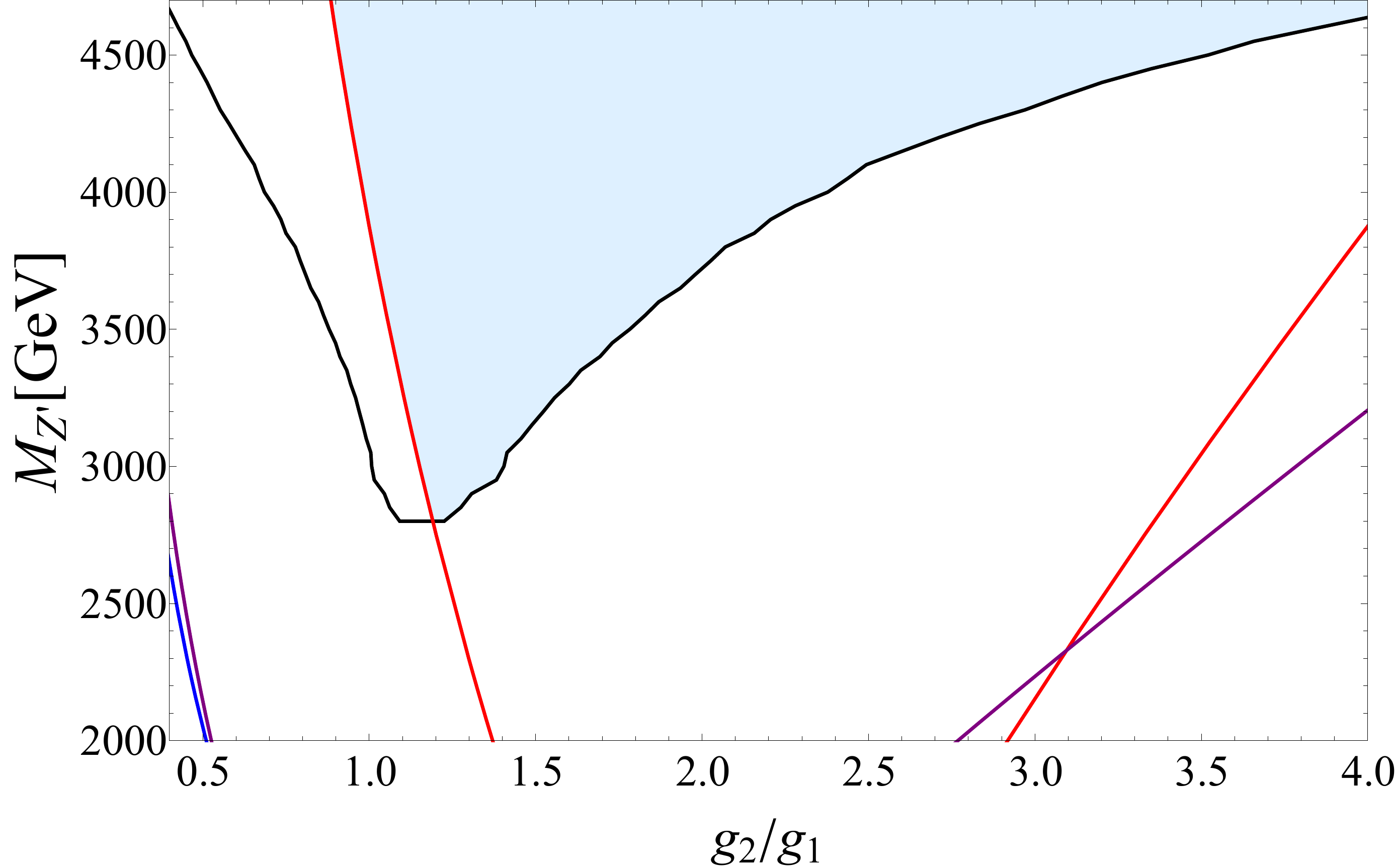}
\end{tabular}
  \caption{The allowed parameter region given by the lightblue region in the $g_2/g_1-M_{Z'}$ plane for several values of $z_1$ and $z_2$, with $m_{\nu'_{3R}}\simeq M_{Z'}/2$ and $m_{\nu'_{1R}}=m_{\nu'_{2R}}=M_{Z'}/3$. The black, blue, red, purple, and orange curves denote the lower bounds obtained from the current LHC limits, precision measurement of the $Z$ decay width, the ambiguity of the $Z$ boson mass, the LEP data, and weak nuclear charge of Cesium, respectively. The region which is below each of these curves is excluded. Top-left panel: $z_1=1/4$ and $z_2=1/8$. Top-right panel: $z_1=1/5$ and $z_2=1/7$. Bottom panel: $z_1=1/5$ and $z_2=1/10$.}\label{APS}
\end{figure}
This figure indicates that the current LHC limits and the ambiguity of the $Z$ boson mass impose the most stringent bounds on the relation between the $Z'$ boson mass and the parameter $t$. Other lower bounds are much weaker than these bounds. In the top-left and bottom panels, we observe that the constraint from the ambiguity of the $Z$ boson mass is stronger(weaker) than that from the current LHC limits if the parameter $t$ is smaller(larger) than about $1.02$ and $1.2$, respectively. In top-right panel, the constraint from the current LHC limits is weaker than that from the ambiguity of the $Z$ boson mass for the regions of the sufficiently small and large $t$, but stronger than for the intermediate region. In addition, we see a weakening in the current LHC limits (as well as in other constraints) appearing around a certain value of $t$ such as $t\approx0.94$ and $t\approx1.46$ for the top-left and top-right panels of Fig. \ref{APS}, respectively. This is due to the behavior of the $Z'$ coupling strength to each fermion which is relatively small around a minimum but becomes larger as going far from this value, as depicted in Fig. \ref{couplings}. And, thus it makes the lower bound of the new gauge boson $Z'$ weaker. Also, we find that the $Z'$ boson mass must not be lower than a minimum value which is about $3$, $0.85$, and $2.8$ TeV corresponding to the top-left, top-right, and bottom panels, respectively.

\section{\label{DM}Dark matter phenomenology}
In this section, we investigate the DM phenomenology where the right-handed neutrino $\nu'_{3R}$ is considered as the DM candidate.\footnote{Other right-handed neutrinos $\nu'_{1,2R}$ can not be the DM candidate. Because of the Yukawa couplings, they can decay into the light SM neutrinos and SM Higgs. In order for the DM candidate, their lifetime must be above the universe age which is around $13.8$ billion years, which corresponds to their decay width have to be below $10^{-42}$ GeV. For the right-handed neutrino masses of a few TeV, the Dirac-Yukawa coupling constants $h^\nu_{ai}$ must be below the order $\mathcal{O}(10^{-23})$, which can not generate the observed neutrino masses.} In our model, the DM $\nu'_{3R}$ communicates with the SM particles through two ways: the Higgs portal and gauge boson portal. Because the DM phenomenology with the Higgs portal has been well investigated in the literature \cite{Okada2010,Okada2012}. Also, the couplings of the new gauge boson $Z'$ to the fermions in our model differ from other models. Therefore, in this work we focus the $\nu'_{3R}$ DM phenomenology corresponding to the gauge boson portal. On the other hand, we assume that the mixing between the SM Higgs and exotic Higgs is negligibly small, e.g. $\lambda_3\ll1$. In addition, as indicated in the previous section that the new gauge boson $Z'$ is heavy enough and hence the mixing between $Z$ and $Z'$ is negligibly small. Thus, with respect to the gauge boson portal the DM $\nu'_{3R}$ communicates with the SM particles mainly through the new gauge boson $Z'$. In the following analysis, we employ $m_{\nu'_{1R}}=m_{\nu'_{2R}}=M_{Z'}/3$.

\subsection{Direct detection}

The effective Lagrangian, which describes the DM scattering on nuclei in the low momentum transfer limit, is obtained by the $t$-channel exchange of the new gauge boson $Z'$ as \cite{Barger2008}
\begin{eqnarray}
\mathcal{L}_{\text{eff}}=\frac{C^{Z'}_{A,\nu'_{R}}}{M^2_{Z'}}(\bar{\nu}'_{3R}\gamma_\mu\gamma^5\nu'_{3R})\bar{q}\gamma^\mu(C^{Z'}_{V,q}+C^{Z'}_{A,q}\gamma^5)q,\label{eff-Lag-DMsc}
\end{eqnarray}
where $C^{Z'}_{A,f}$ is the axial-vector coupling of $Z'$ to the fermion $f$. Note that, because the DM in this work is the self-conjugate field, there is no vector-vector coupling in the above effective Lagrangian. The Lagrangian (\ref{eff-Lag-DMsc}) leads to the spin-dependent (SD) scattering and in the case of that the nucleus target is neutron the cross-section is given as \cite{Barger2008}
\begin{eqnarray}
\sigma_{\text{SD}}=(2!)^2\frac{3m^2_{\text{DM-n}}}{\pi M^4_{Z'}}\left(C^{Z'}_{A,\nu'_{R}}\right)^2\left[C^{Z'}_{A,u}\lambda^{(\text{n})}_u+C^{Z'}_{A,d}\left(\lambda^{(\text{n})}_d+\lambda^{(\text{n})}_s\right)\right]^2
\end{eqnarray}
where 
\begin{eqnarray}
m_{\text{DM-n}}=\frac{m_{\nu'_{3R}}m_{\text{n}}}{m_{\nu'_{3R}}+m_{\text{n}}},
\end{eqnarray}
with $m_{\text{n}}$ to be the mass of proton and $\lambda^{(\text{n})}_q$ are the fractional quark-spin coefficients of proton given explicitly as \cite{Chiang2012}, $\lambda^{(\text{n})}_u=-0.42$, $\lambda^{(\text{n})}_d=0.85$, $\lambda^{(\text{n})}_s=-0.08$.

In Fig. \ref{DM0}, we show the DM-neutron cross-section as a function of $M_{Z'}$ for various values of $t$ and $z_{1,2}$ along with the $90\%$ C.L. upper limits from XENON100 experiment \cite{XENON100}. From this figure, it is obvious that the DM-neutron cross-section associated with the allowed parameter region obtained in the previous section is below the XENON100 $90\%$ C.L. upper limits.
\begin{figure}[t]
 \centering
\begin{tabular}{cc}
\includegraphics[width=0.47 \textwidth]{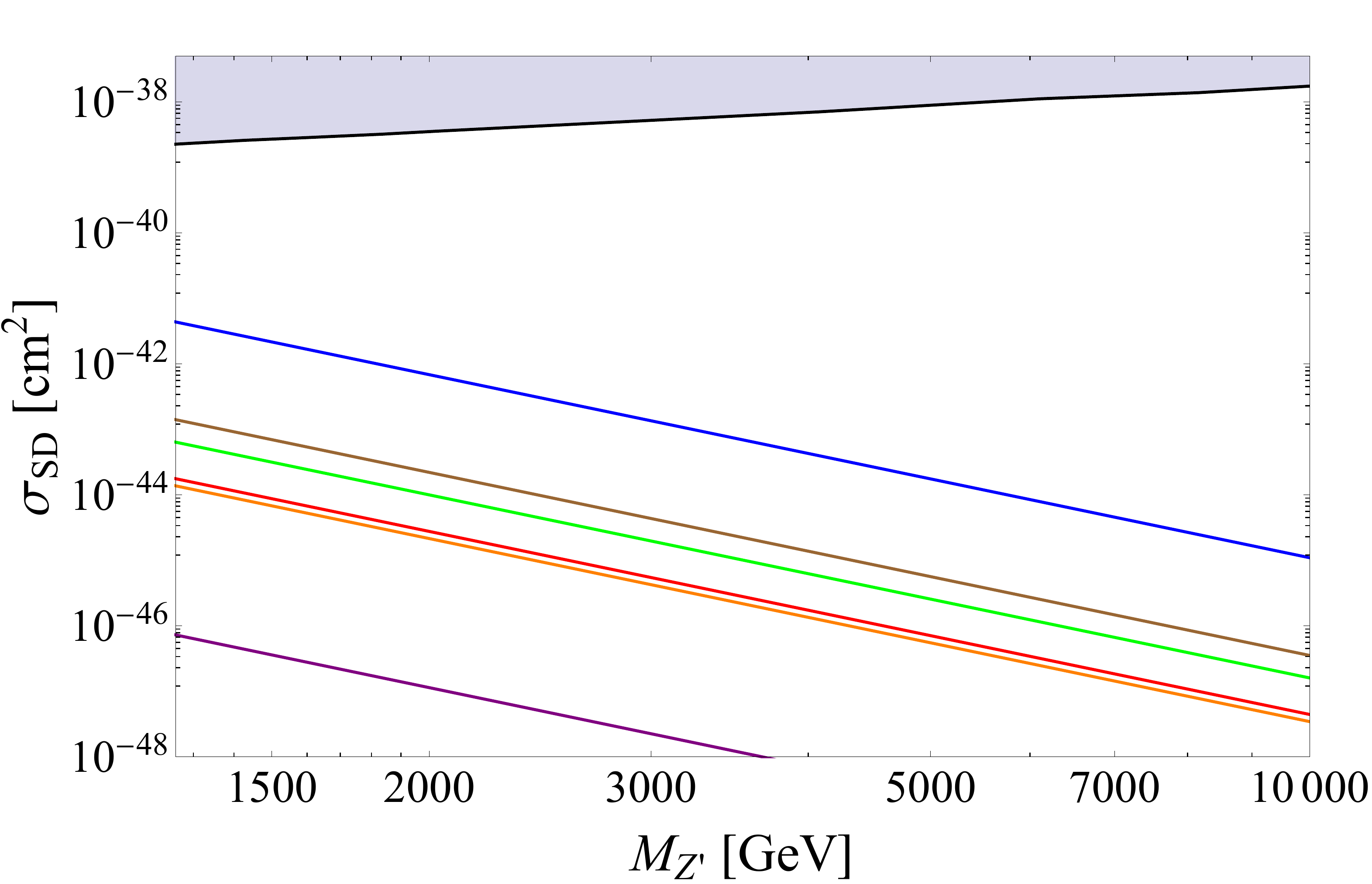}
\hspace*{0.05\textwidth}
\includegraphics[width=0.47 \textwidth]{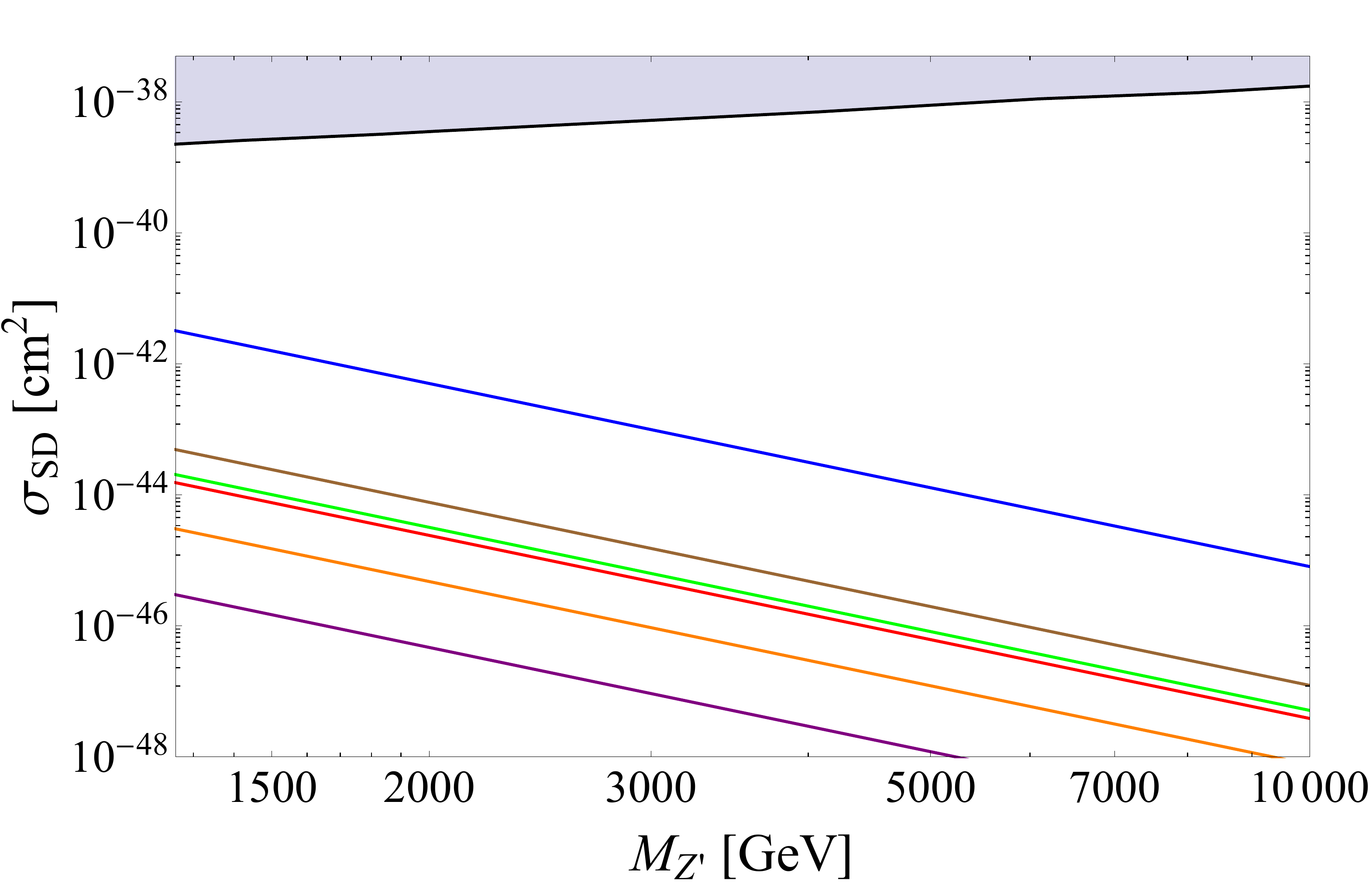}\\
\end{tabular}
  \caption{The DM-neutron cross-section as a function of $M_{Z'}$ with $m_{\nu'_{3R}}\simeq M_{Z'}/2$. The blue, red, purple, orange, green, and brown curves correspond to $t=0.2$, $0.9$, $1.8$, $2.6$, $3.4$, and $4.0$, respectively. The black curve refers to the $90\%$ C.L. upper limits from XENON100 experiment. Left panel: $z_1=1/4$ and $z_2=1/8$. Right panel: $z_1=1/5$ and $z_2=1/10$.}\label{DM0}
\end{figure}

\subsection{Indirect detection}
During the early universe with the efficiently high temperature, the DM particles are in the thermal equilibrium with the thermal bath of the SM particles. However, when the temperature drops below the DM mass, the rate of the annihilation per DM particle becomes smaller than the rate of the Hubble expansion and hence it would lead to the freeze-out of the DM species. The DM relic abundance is observed by the Planck
satellite experiments as, $\Omega_{\text{DM}}h^2=0.1198\pm0.0015$ \cite{Ade2016}. 

The DM relic density is given by the following analytic approximation \cite{Gondolo1991}
\begin{eqnarray}
\Omega_{\text{DM}}h^2=\frac{1.07\times10^9\ \ \text{GeV}^{-1}}{g^{1/2}_*J(x_f)M_{\text{Pl}}},
\end{eqnarray}
where $g_*$ is the effective total number of degrees of freedom for the particles at the time of freeze-out (for the following analysis we employ $g_*$=106.75 which is for the SM particles), $M_{\text{Pl}}=1.22\times10^{19}$ GeV is the Planck scale. The function $J(x_f)$ is given by
\begin{eqnarray}
J(x_f)&=&\int^{\infty}_{x_f}\frac{\langle\sigma v\rangle(x)}{x^2}dx,\nonumber\\
\langle\sigma v\rangle(x)&=&\frac{x}{8m^5_{\nu'_{3R}}[K_2(x)]^2}\int^{\infty}_{4m^2_{\nu'_{3R}}}\sigma\times(s-4m^2_{\nu'_{3R}})\sqrt{s}K_1\left(\frac{x\sqrt{s}}{m_{\nu'_{3R}}}\right)ds,\label{totcroos}
\end{eqnarray}
where $\sigma$ is the total cross section of the DM pair annihilation into the SM particles as well as other exotic particles and the freeze-out parameter $x_f$ is computed by the following equation
\begin{eqnarray}
x_f=\ln\left(\frac{0.038gM_{\text{Pl}}m_{\text{DM}}\langle\sigma v\rangle(x_f)}{\left(g_*x_f\right)^{1/2}}\right),
\end{eqnarray}
which is about $25$ for the DM mass at a few TeV, with $g=2$ to be the number of degrees of freedom for the DM particle in consideration. In this work we are only interested the DM mass satisfying
\begin{eqnarray}
m_{\nu'_{1R}},m_{\nu'_{1R}}<m_{\nu'_{3R}}&<&\frac{M_{Z'}}{2}<\frac{m_{h_2}}{2},\label{asumpt}
\end{eqnarray}
where $h_2$ refers to the exotic Higgs. This means that the two-body annihilation channels of the DM pair through the $Z'$ portal are
\begin{eqnarray}
\bar{\nu}'_{3R}\nu'_{3R}\rightarrow\bar{f}f,\bar{\nu}'_{iR}\nu'_{iR},W^+W^-,Zh_1,
\end{eqnarray}
where $f$ represents the SM fermions and $i=1,2$. Due to the mixing angle between $Z$ and $Z'$ to be negligibly small, the contributions from the DM pair annihilation into $W^+W^-$ and $Zh_1$ are ignored. As a result, the total cross section of the DM pair annihilation is approximately given by
\begin{eqnarray}
\sigma &=&\sigma_{\text{SM}}+\sum_{i=1,2}\sigma_{\nu'_{iR}},\nonumber\\
\sigma_{\text{SM}}&=&\frac{1}{\pi}\left(\frac{gt_W}{12t}\right)^4\left[6(1+t^2)(3z_1-4z_2)\right]^2\frac{\sqrt{s\left(s-4m^2_{\nu'_{3R}}\right)}}{\left(s-M^2_{Z'}\right)^2+M^2_{Z'}\Gamma^2_{Z'}}\nonumber\\
&&\times\left[F_1(t,z_1,z_2)+F_2(t,z_1,z_2)\left(1+\frac{2m^2_t}{s}\right)\sqrt{1-\frac{4m^2_t}{s}}\right],\nonumber\\
\sigma_{\nu'_{iR}}&=&\frac{1}{48\pi}\left[\frac{gt_W(1+t^2)(3z_1-4z_2)}{t}\right]^4\frac{1}{s\left[\left(s-M^2_{Z'}\right)^2+M^2_{Z'}\Gamma^2_{Z'}\right]}\sqrt{\frac{s-4m^2_{\nu'_{iR}}}{s-4m^2_{\nu'_{3R}}}}\nonumber\\&&\times\left[\left(s-4m^2_{\nu'_{3R}}\right)\left(s-4m^2_{\nu'_{iR}}\right)+\frac{12m^2_{\nu'_{3R}}m^2_{\nu'_{iR}}}{M^4_{Z'}}\left(s-M^2_{Z'}\right)^2\right]
\end{eqnarray}
where $\sigma_{\text{SM}}$ and $\sigma_{\nu'_{iR}}$ describe the DM pair annihilation into the SM fermions and the heavy neutrino pair $\bar{\nu}'_{iR}\nu'_{iR}$, respectively, and
\begin{eqnarray}
F_1(t,z_1,z_2)&=&103+12(1+t^2)\left[(23z_2-43z_1)+3(1+t^2)(37z^2_1+39z^2_2-70z_1z_2)\right],\nonumber\\
F_2(t,z_1,z_2)&=&17+12(1+t^2)\left[(z_2-5z_1)+3(1+t^2)(2z^2_1+z^2_2-2z_1z_2)\right].
\end{eqnarray}
The relic density of the DM $\nu'_{3R}$ depends on five parameters which are $t$, $z_{1,2}$, $m_{\nu'_{3R}}$, and $M_{Z'}$.

In Fig. \ref{DM1}, we show the prediction of our model for the DM relic abundance as a function of the DM mass for various values of $t$ at which the $Z'$ boson mass and $z_{1,2}$ are kept fixed all. In the left panel, the blue, red, and purple curves correspond to $t=1.0$, $1.2$, and $1.4$, respectively, and we have fixed $2z_1=z_2=1/8$ and $M_{Z'}=4$ TeV. From this panel and the top-left panel of Fig. \ref{APS}, we can see that for $M_{Z'}=4$ TeV most of values of $t$ constrained in the previous section can lead to the range of the observed DM relic abundance $0.1183\leq\Omega_{\text{DM}}h^2\leq0.1213$. In the right panel, the blue, red, purple, and orange curves correspond to $t=1.3$, $1.415$, and $1.5$, and $1.6$, respectively, and we have fixed $2z_1=z_2=1/10$ and $M_{Z'}=3.5$ TeV. This panel indicates that the value of $t$ should be above the lower bound $1.415$ to lead to the range of the observed DM relic abundance. In addition, Fig. \ref{DM1} shows that, in order to achieve the observed DM relic abundance, the DM mass should be around $M_{Z'}/2$ corresponding to the $Z'$ boson resonance.
\begin{figure}[t]
 \centering
\begin{tabular}{cc}
\includegraphics[width=0.47 \textwidth]{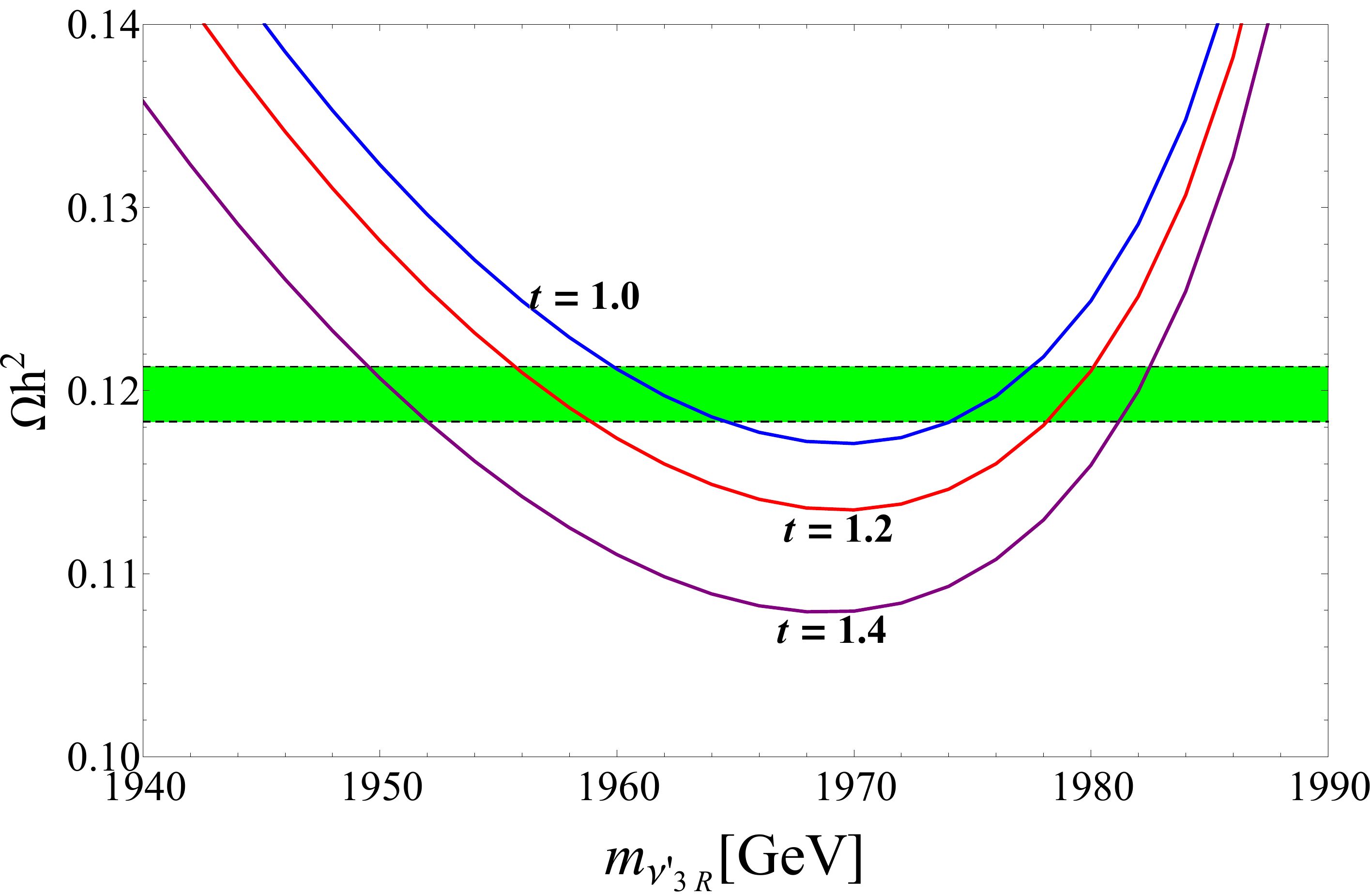}
\hspace*{0.05\textwidth}
\includegraphics[width=0.47 \textwidth]{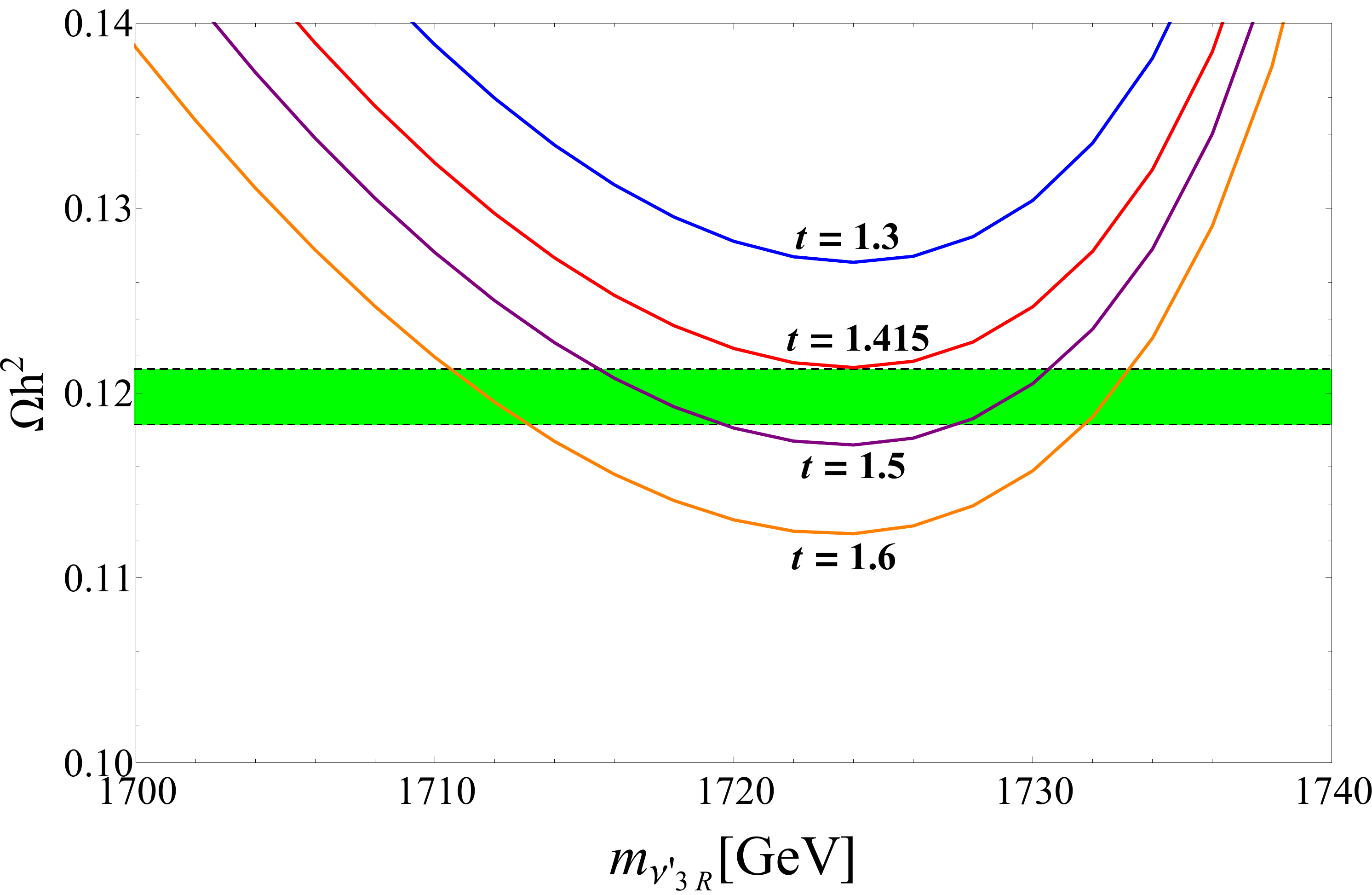}\\
\end{tabular}
  \caption{The DM relic abundance as a function of the DM mass for various values of $t$. The green band refers to the range of the observed DM relic abundance $0.1183\leq\Omega_{\text{DM}}h^2\leq0.1213$. Left panel: $2z_1=z_2=1/8$ and $M_{Z'}=4$ TeV. Right panel: $2z_1=z_2=1/10$ and $M_{Z'}=3.5$ TeV.}\label{DM1}
\end{figure}

In Fig. \ref{DM2}, we show the prediction of our model for the DM relic abundance as a function of $t$ for various values of the $Z'$ gauge boson mass with $m_{\nu'_{3R}}\simeq M_{Z'}/2$.
\begin{figure}[t]
 \centering
\begin{tabular}{cc}
\includegraphics[width=0.47 \textwidth]{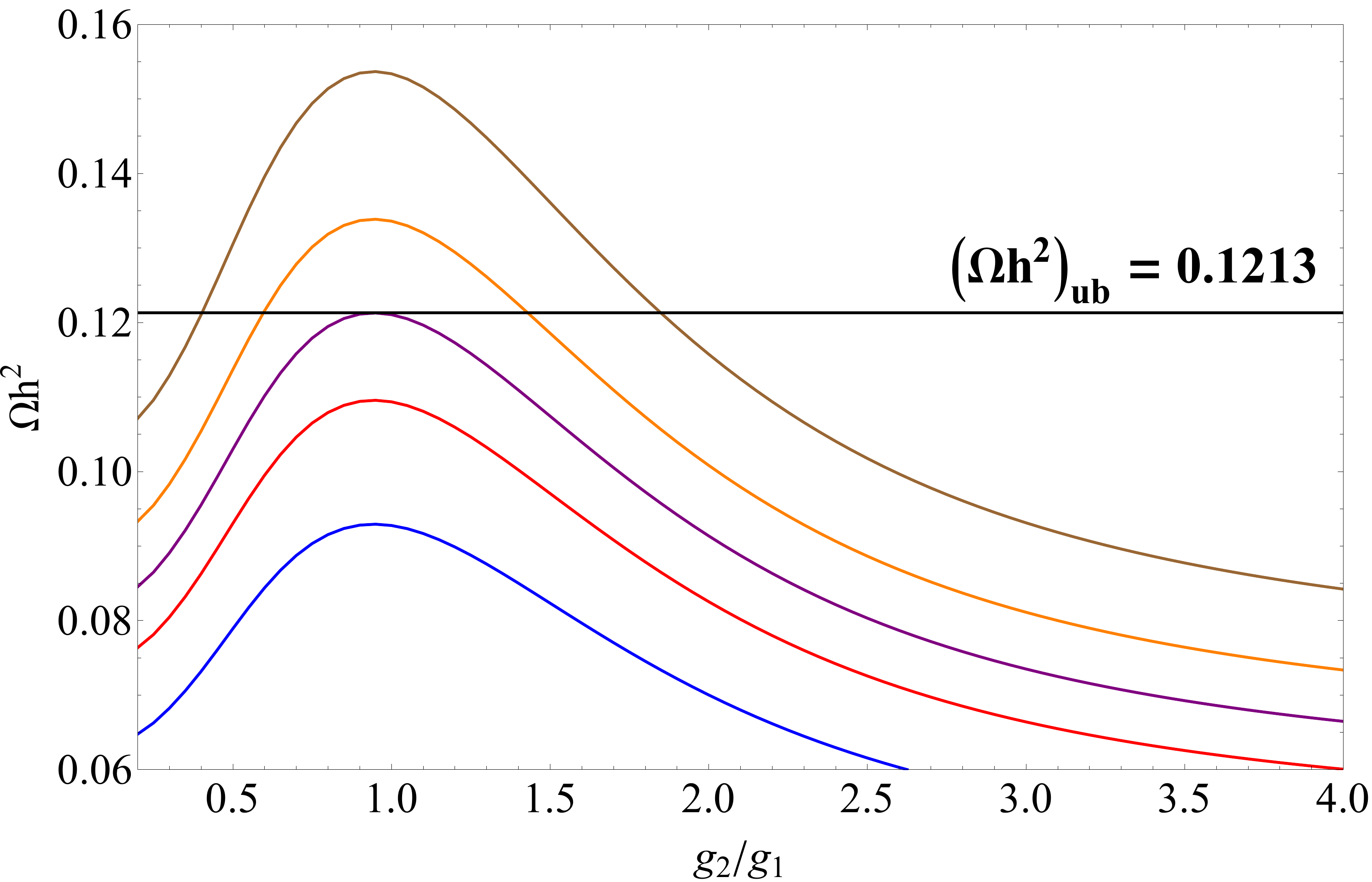}
\hspace*{0.05\textwidth}
\includegraphics[width=0.47 \textwidth]{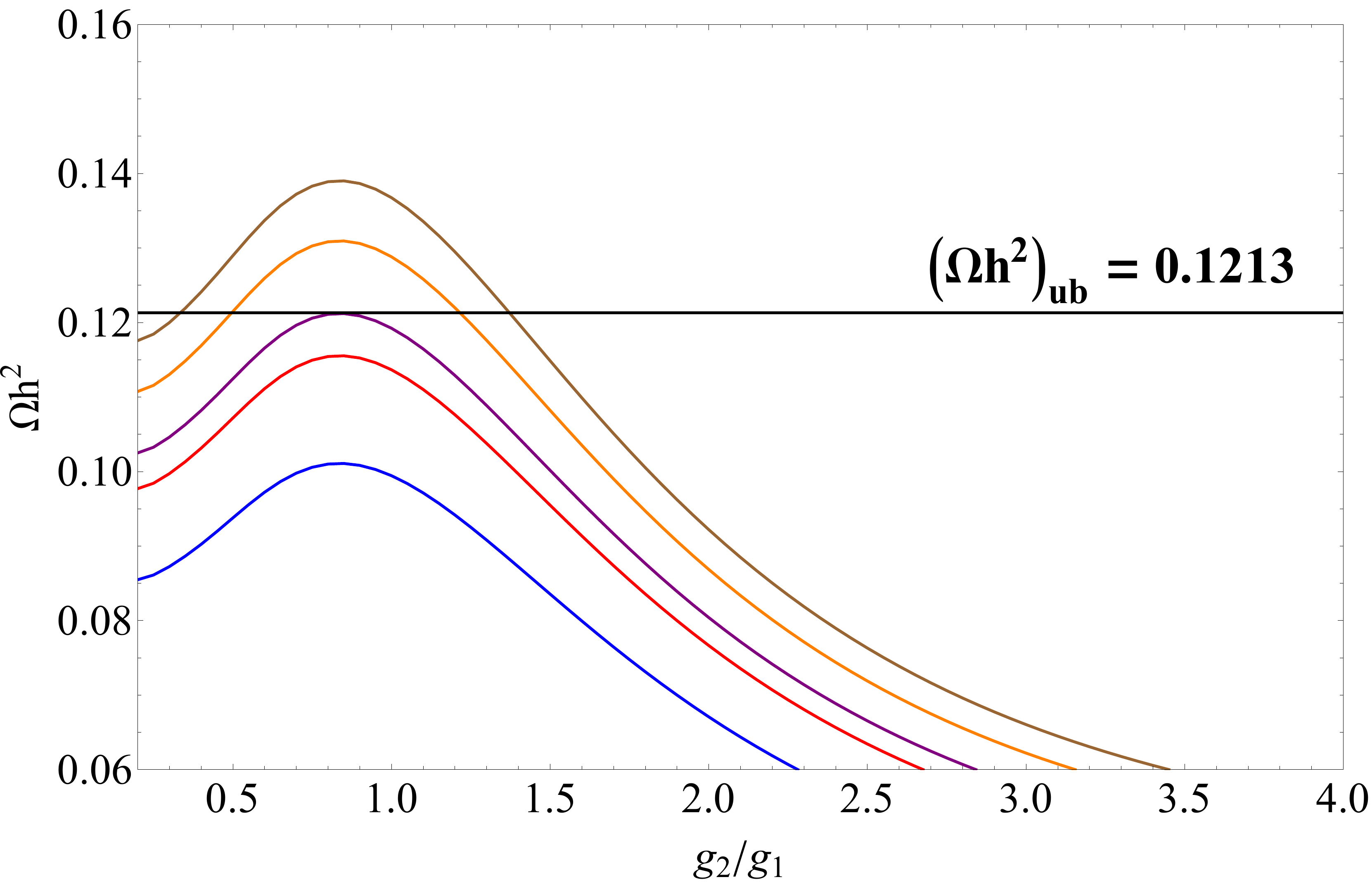}\\
\end{tabular}
  \caption{The DM relic abundance as a function of $g_2/g_1$ for various values of the $Z'$ boson mass. In the left(right) panel, the blue, red, purple, orange, and brown curevs correspond to $M_{Z'}=3.5(2.9)$, $3.8(3.1)$, $3.998(3.175)$, $4.2(3.3)$, and $4.5(3.4)$ TeV, respectively. The black horizontal line represents the upper bound of the DM relic abundance. Top-left panel: $z_1=1/4$ and $z_2=1/8$. Top-right panel: $z_1=1/5$ and $z_2=1/10$. }\label{DM2}
\end{figure}
From this figure, we find that if the $Z'$ gauge boson mass is below a certain value which is about $3998(3175)$ GeV for the left(right) panel, almost values of $t$ satisfy the upper bound of the DM relic abundance $\left(\Omega_{\text{DM}}h^2\right)_{\text{ub}}=0.1213$. On the contrary, only the sufficiently small or large $t$ values satisfy this upper bound.

In Fig. \ref{DM3}, we show the allowed parameter region in the $t-M_{Z'}$ plane which is consistent with the constraints discussed in the previous section and the observed DM relic abundance.
\begin{figure}[t]
 \centering
\begin{tabular}{cc}
\includegraphics[width=0.47 \textwidth]{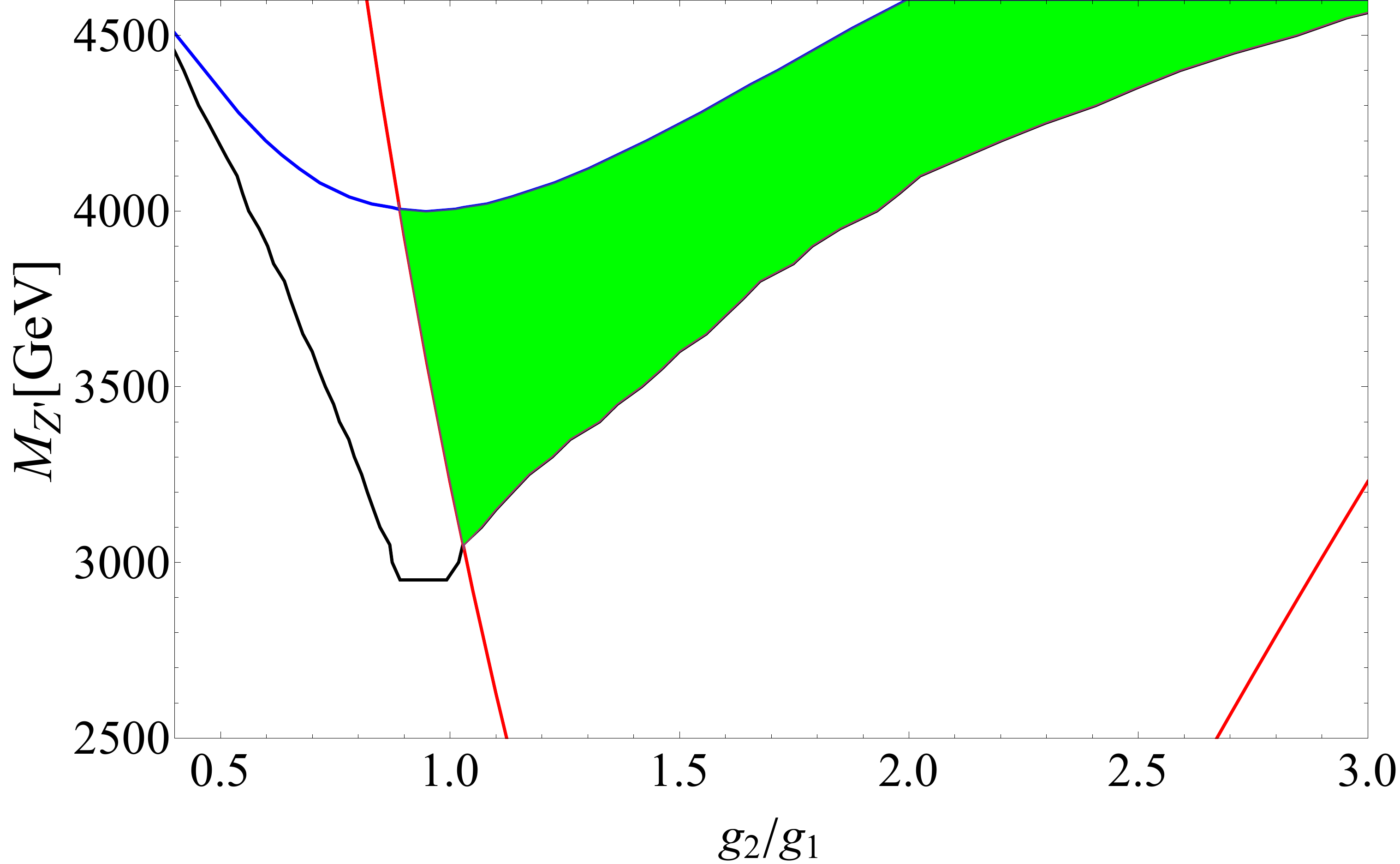}
\hspace*{0.05\textwidth}
\includegraphics[width=0.47 \textwidth]{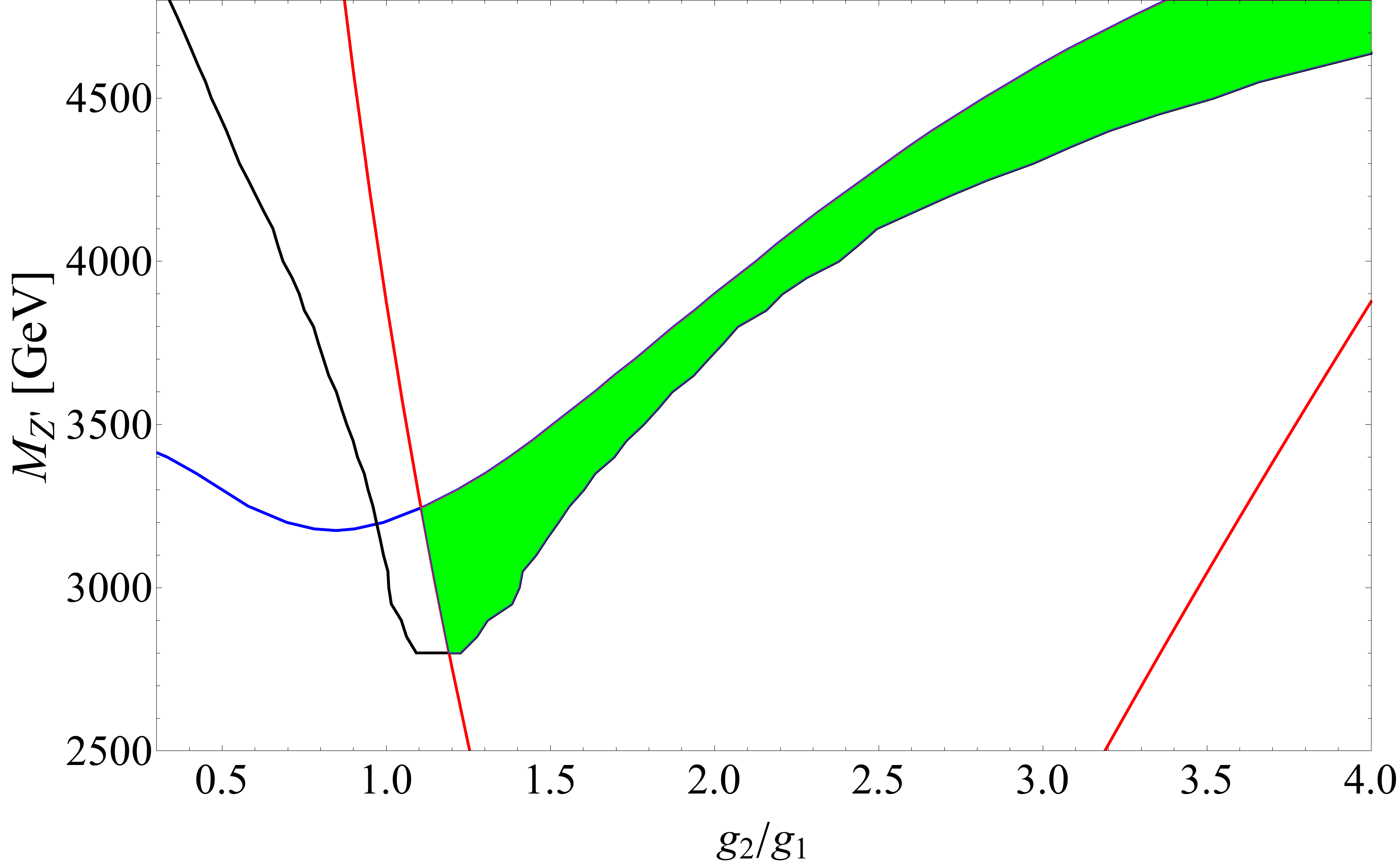}\\
\end{tabular}
  \caption{Allowed parameter region given by the green region in the $g_2/g_1-M_{Z'}$ plane for several values of $z_1$ and $z_2$, with $m_{\nu'_{3R}}\simeq M_{Z'}/2$. The region above the blue curve is excluded by the upper bound of the DM relic abundance, whereas the regions below the black and red curves are excluded by the current LHC limits and the ambiguity of the $Z$ boson mass, respectively. Top-left panel: $z_1=1/4$ and $z_2=1/8$. Top-right panel: $z_1=1/5$ and $z_2=1/10$.}\label{DM3}
\end{figure}
The blue curve represents the upper bound on $M_{Z'}$ as a function of $t$ obtained from the constraint which the relic density of the DM candidate $\nu'_{3R}$ must not be larger than the upper bound of the DM relic abundance. Whereas, the black and red curves represent the lower bounds on $M_{Z'}$ as the functions of $t$, which are obtained from the current LHC limits and the ambiguity of the $Z$ gauge boson mass, respectively. The green region refers to the allowed parameter region after combining these three constraints. It indicates lower bounds for the parameter $t$ and the $Z'$ gauge boson mass, $t\gtrsim0.9(1.1)$ and $M_{Z'}\gtrsim3.0(2.8)$ TeV for $2z_1=z_2=1/8$($2z_1=z_2=1/10$).

\section{\label{conclu}Conclusion}

In this work, we have constructed a general flavor-independent model in a minimal way with the symmetry $SU(3)_C\times SU(2)_L\times U(1)_{Y'}\times U(1)_X\times Z_2$ and the charge operator identified in terms of both $U(1)_{Y'}$ and $U(1)_X$ charges. We have determined the gauge charges of the fields relying on the conditions of the anomaly cancellation and the gauge invariance of the Yukawa couplings. The smallness of active neutrino masses is explained through Type-I seesaw mechanism only with two heavy right-handed neutrinos whose Majorana masses are determined by the $U(1)_{Y'}\times U(1)_X$ symmetry breaking scale. 
We have obtained the physical states of the neutral gauge bosons and their couplings to the fermions. We have studied the constraints on the mass and gauge coupling of the new gauge boson coming from various current experiments at which the current LHC limits and the ambiguity of the Standard Model neutral gauge boson mass impose the most stringent bounds. Finally, we have investigated the phenomenology of the $Z_2$-odd right-handed neutrino dark matter. We have identified the allowed parameter space of the model which is consistent with the upper bound on the dark matter relic abundance and other bounds.

\section*{Acknowledgements}
I am grateful Theoretical Physics Group at IFIRSE for the warm hospitality during my visit. I would like to thank Dr. LE Duc Ninh, Dr. DAO Thi Nhung, and members in their group for the useful discussions.

\end{document}